
 
\ifx\tenpoint\undefined\let\loadedfrommacro=Y%
\ifx\loadedfrommacro Y\else
         \message{10point.TeX must be loaded from a macro package.}
         \message{Input terminated.}
          \fi
 
\font\tencsc=cmcsc10
 
\newfam\scfam
 
\def\tenpoint{\def\rm{\fam0\tenrm}
    \textfont0=\tenrm  \scriptfont0=\sevenrm  \scriptscriptfont0=\fiverm
    \textfont1=\teni   \scriptfont1=\seveni   \scriptscriptfont1=\fivei
    \textfont2=\tensy  \scriptfont2=\sevensy  \scriptscriptfont2=\fivesy
    \textfont3=\tenex  \scriptfont3=\tenex    \scriptscriptfont3=\tenex
    \textfont\itfam=\tenit   \def\it{\fam\itfam\tenit}%
    \textfont\slfam=\tensl   \def\sl{\fam\slfam\tensl}%
    \textfont\ttfam=\tentt   \def\tt{\fam\ttfam\tentt}%
    \textfont\bffam=\tenbf   \scriptfont\bffam=\sevenbf
    \scriptscriptfont\bffam=\fivebf  \def\bf{\fam\bffam\tenbf}%
    \textfont\scfam=\tencsc  \def\sc{\fam\scfam\tencsc}%
    \normalbaselineskip=12pt
    \setbox\strutbox=\hbox{\vrule height8.5pt depth 3.5pt width0pt}%
    \normalbaselines\rm}

         \let\loadedfrommacro=N\fi
\font\ninerm=cmr9            \font\sixrm=cmr6
\font\ninei=cmmi9            \font\sixi=cmmi6
\font\ninesy=cmsy9           \font\sixsy=cmsy6
\font\ninebf=cmbx9           \font\sixbf=cmbx6
\font\ninesl=cmsl9           \font\ninett=cmtt9      \font\nineit=cmti9
\font\ninecsc=cmcsc10
\font\ninebfit=cmbxti10 at 9pt
\ifx\ninepoint\undefined
   \def\ninepoint{\def\rm{\fam0\ninerm}
       \textfont0=\ninerm  \scriptfont0=\sixrm  \scriptscriptfont0=\fiverm
       \textfont1=\ninei   \scriptfont1=\sixi   \scriptscriptfont1=\fivei
       \textfont2=\ninesy  \scriptfont2=\sixsy  \scriptscriptfont2=\fivesy
       \textfont3=\tenex   \scriptfont3=\tenex  \scriptscriptfont3=\tenex
       \def\bfit{\ninebfit}%
       \textfont\itfam=\nineit   \def\it{\fam\itfam\nineit}%
       \textfont\slfam=\ninesl   \def\sl{\fam\slfam\ninesl}%
       \textfont\ttfam=\ninett   \def\tt{\fam\ttfam\ninett}%
       \textfont\bffam=\ninebf   \scriptfont\bffam=\sixbf
        \scriptscriptfont\bffam=\fivebf   \def\bf{\fam\bffam\ninebf}%
       \textfont\scfam=\ninecsc  \def\sc{\fam\scfam\ninecsc}%
       \normalbaselineskip=11pt%
       \setbox\strutbox=\hbox{\vrule height8pt depth3pt width0pt}%
       \normalbaselines\rm}%
   \fi

 
\ifx\tenpoint\undefined\let\loadedfrommacro=Y
\ifx\loadedfrommacro Y\else
         \message{10point.TeX must be loaded from a macro package.}
         \message{Input terminated.}
          \fi
 
\font\tencsc=cmcsc10
 
\newfam\scfam
 
\def\tenpoint{\def\rm{\fam0\tenrm}
    \textfont0=\tenrm  \scriptfont0=\sevenrm  \scriptscriptfont0=\fiverm
    \textfont1=\teni   \scriptfont1=\seveni   \scriptscriptfont1=\fivei
    \textfont2=\tensy  \scriptfont2=\sevensy  \scriptscriptfont2=\fivesy
    \textfont3=\tenex  \scriptfont3=\tenex    \scriptscriptfont3=\tenex
    \textfont\itfam=\tenit   \def\it{\fam\itfam\tenit}%
    \textfont\slfam=\tensl   \def\sl{\fam\slfam\tensl}%
    \textfont\ttfam=\tentt   \def\tt{\fam\ttfam\tentt}%
    \textfont\bffam=\tenbf   \scriptfont\bffam=\sevenbf
    \scriptscriptfont\bffam=\fivebf  \def\bf{\fam\bffam\tenbf}%
    \textfont\scfam=\tencsc  \def\sc{\fam\scfam\tencsc}%
    \normalbaselineskip=12pt
    \setbox\strutbox=\hbox{\vrule height8.5pt depth 3.5pt width0pt}%
    \normalbaselines\rm}

         \let\loadedfrommacro=N\fi
 
\font\elevenrm=cmr10    scaled \magstephalf
\font\eleveni=cmmi10    scaled \magstephalf
\font\elevensy=cmsy10   scaled \magstephalf
\font\elevenex=cmex10   scaled \magstephalf
\font\elevenbf=cmbx10   scaled \magstephalf
\font\elevensl=cmsl10   scaled \magstephalf
\font\eleventt=cmtt10   scaled \magstephalf
\font\elevenit=cmti10   scaled \magstephalf
\font\elevencsc=cmcsc10 scaled \magstephalf
 
\font\eightrm=cmr8           \font\sixrm=cmr6
\font\eighti=cmmi8           \font\sixi=cmmi6
\font\eightsy=cmsy8          \font\sixsy=cmsy6
\font\eightbf=cmbx8          \font\sixbf=cmbx6
 
\ifx\elevenpoint\undefined
   \def\elevenpoint{\def\rm{\fam0\elevenrm}
       \textfont0=\elevenrm \scriptfont0=\eightrm \scriptscriptfont0=\sixrm
       \textfont1=\eleveni  \scriptfont1=\eighti  \scriptscriptfont1=\sixi
       \textfont2=\elevensy \scriptfont2=\eightsy \scriptscriptfont2=\sixsy
       \textfont3=\elevenex \scriptfont3=\elevenex\scriptscriptfont3=\elevenex
       \textfont\itfam=\elevenit  \def\it{\fam\itfam\elevenit}%
       \textfont\slfam=\elevensl  \def\sl{\fam\slfam\elevensl}%
       \textfont\ttfam=\eleventt  \def\tt{\fam\ttfam\eleventt}%
       \textfont\bffam=\elevenbf  \scriptfont\bffam=\eightbf
        \scriptscriptfont\bffam=\sixbf  \def\bf{\fam\bffam\elevenbf}%
       \textfont\scfam=\elevencsc \def\sc{\fam\scfam\elevencsc}%
       \normalbaselineskip=14pt
       \setbox\strutbox=\hbox{\vrule height9pt depth4pt width0pt}%
       \normalbaselines\rm}
   \fi

\input eplain
\input epsf
\input colordvi

\font\bigbf=cmbx10 at 14pt
\font\tenbfit=cmbxti10
\def\titlebf{\bigbf}

\def\by#1{\noindent {\it #1}}
\def\runningtitle#1{\headline={\tenit \hss
\ifnum\pageno>1 #1%
\fi}}
\def\abstract{\begingroup \bigskip\bigskip
\hrule\medskip\tenpoint \noindent\ignorespaces {\bf Abstract.}~}
\def\endabstract{\medskip\endgroup\hrule\bigskip}

\def\title#1{%
\begingroup
\parindent=0pt \baselineskip=17pt {\titlebf #1}\par 
\par\endgroup}

\bigskip\bigskip

\newcount\notenum \notenum=0
\def\note{\global\advance\notenum by 1%
\footnote{\the\notenum}}

\newcount \sectionnum \sectionnum=0
\def\section#1 \par{\bigbreak\bigskip
\global\advance\sectionnum by 1%
\leftline{\bigbf \llap{\the\sectionnum\hskip1.5ex}#1}%
\nobreak\medskip
\noindent\ignorespaces}

\def\lsection#1#2 \par{\section{#2} \par
\definexref{#1}{\the\sectionnum}{section}}

\def\beginquote{\begingroup\par\vskip-\parskip\smallskip
\tenpoint
\leftskip=\a \rightskip=\a
\noindent \ignorespaces}

\def\endquote{\par\vskip-\parskip\smallskip\endgroup\noindent\ignorespaces}

\def\refdefs{\frenchspacing
\advance\baselineskip by -2pt%
\parskip=4pt plus 0.5pt minus 0.5pt%
\advance\rightskip by 0pt plus 1em%
\def\1{\discretionary{}{}{}}%
\parindent=0pt\relax
\everypar={\hangindent=\a \leftskip=0pt \hangafter=1\relax}%
\def\beginquote{\begingroup \vskip1.5pt \everypar={}%
\tenpoint
\leftskip=2\a \rightskip=2\a
\noindent \ignorespaces}%
\def\hb{\hfil\break}
\def\endquote{\vskip1.5pt\endgroup}%
\def\ajp{\j{American Journal of Physics}}%
\def\pt{\j{Physics Teacher}}%
\def\t##1{{\it ##1}}%
\def\nj##1##2{{\bf ##1}:$\,$##2}%
\def\j##1##2##3{{\it ##1\/} {\bf ##2}:$\,$##3}}

\newdimen\a \a=\parindent

\newcount\tablenum \tablenum=0
\def\tableword{Table}

\newbox\caption
\def\begintable#1#2{\midinsert
\global\advance\tablenum by 1%
\definexref{#1}{\the\tablenum}{table}%
\tenpoint
\def\1{\noalign{\smallskip \hrule \vskip1pt\hrule \medskip}}%
\def\2{\noalign{\smallskip\hrule\medskip}}
\def\3{\noalign{\medskip \hrule \vskip1pt\hrule \smallskip}}%
\everycr={\noalign{\medskip}}%
\setbox\caption=\hbox{{\tenbfit \tableword~\the\tablenum.}~\it #2}%
\ifdim\wd\caption>5truein%
\setbox\caption=\vbox{\hsize=5truein%
\unhbox\caption}%
\else
\setbox\caption=\vbox{\centerline{\box\caption}}%
\fi
\noindent \box\caption \par
\vbox\bgroup}

\def\endtable{\egroup\endinsert}

\def\rawmps#1{\vbox{\epsfbox{#1}}}

\def\rawurlaux#1#2{\ldefs
\tdefs
\textRed
$\langle$#2$\rangle$%
\textBlack
\egroup}
\def\urlaux#1{\rawurlaux{#1}{#1}}

\begingroup
\catcode`_=13 
\catcode`\&=13
\catcode`\$=13
\catcode`-=13
\catcode`/=13
\gdef\fixu{\def_{\leavevmode \kern.06em \vbox{\hrule width.3em}}}
\catcode`.=13
\gdef\tdefs{\fixu
\def&{\char38}%
\def.{\discretionary{\char46}{}{\char46}}%
\def-{\discretionary{\char45}{}{\char45}}%
\def/{\discretionary{\char47}{}{\char47}}%
\def~{\char126}}

\gdef\ldefs{\chardef\_=`\_ \let_=\_%
\let&=\&%
\let$=\$%
\chardef\~=`\~%
\let~=\~%
\chardef\.=`.%
\let.=\.%
\chardef\-=`-%
\let-=\-%
\chardef\/=`/%
\let/=\/%
}

\gdef\rawurl{\bgroup\catcode`~=13 \catcode`_=13 \catcode`\&=13 \catcode`\$=13%
\rawurlaux}

\gdef\url{\bgroup\catcode`~=13 \catcode`_=13 \catcode`\&=13 \catcode`\$=13%
\catcode`.=13 \catcode`=13 \catcode`/=13%
\urlaux}

\endgroup

\newcount\fignum
\fignum=0

\def\figlegend#1#2{\global\advance\fignum by 1%
\definexref{#1}{\the\fignum}{figure}%
\medskip
\vbox{\baselineskip=11pt%
\leftskip=\a \rightskip=\a
\noindent {\tenbfit Figure~\the\fignum.}~\tenit #2}}

\def\mpsfig#1#2#3{\midinsert
\centerline{\rawmps{#2}}
\figlegend{#1}{#3}\endinsert}

\newbox\ftnumbox
\catcode`@=11
\topskip 10\p@ plus10\p@ \r@ggedbottomtrue
\def\footnote#1{\let\@sf\empty 
  \ifhmode\edef\@sf{\spacefactor\the\spacefactor}\/\fi
  $^{#1}$%
  \@sf\vfootnote{#1.}}
\def\vfootnote#1{\insert\footins\bgroup
  \ninepoint \baselineskip=9pt%
  \global\setbox\ftnumbox=\hbox{#1\enspace}%
  \hangindent=\wd\ftnumbox \hangafter=1%
  \interlinepenalty\interfootnotelinepenalty
  \splittopskip\ht\strutbox 
  \splitmaxdepth\dp\strutbox \floatingpenalty\@MM
  \leftskip\z@skip \rightskip\z@skip \spaceskip\z@skip \xspaceskip\z@skip
  \noindent{\copy\ftnumbox}\footstrut\futurelet\next\fo@t}
\def\f@t#1{#1\@foot}
\catcode`@=12 

\runningtitle{Benezet--Berman mathematics teaching experiment}

\elevenpoint

\title{Is it time for a science counterpart of the Benezet--Berman
mathematics teaching experiment of the 1930's?}
\medskip
\parindent=0pt
\by{Sanjoy Mahajan},\note{Supported by the Gatsby Charitable
Foundation.}
University of Cambridge \url{http://www.inference.phy.cam.ac.uk/sanjoy}
\par
\by{Richard R. Hake},\note{Partially supported by NSF Grant DUE/MDR-9253965}
Indiana University \url{http://www.physics.indiana.edu/~hake}

\medskip
\noindent Based on an invited poster presented at the {\it Physics Education
Research Conference 2000: Teacher Education\/}
\url{http://www.sci.ccny.cuny.edu/~rstein/perc2000.htm}

\abstract
Should teachers concentrate on critical thinking, estimation,
measurement, and graphing rather than college-clone algorithmic
physics in grades K--12?\note{For non-US readers: Children begin
kindergarten (grade K) roughly at age 5 and finish the highest grade
(12)
roughly at age 18.}  Thus far physics education research offers little
substantive guidance.  Mathematics education research addressed the
mathematics analogue of this question in the 1930's. Students in
Manchester, New Hampshire were not subjected to arithmetic algorithms
until grade 6.  In earlier grades they read, invented, and discussed
stories and problems; estimated lengths, heights, and areas; and
enjoyed finding and interpreting numbers relevant to their lives.  In
grade 6, with 4 months of formal training, they caught up to the
regular students in algorithmic ability, and were far ahead in general
numeracy and in the verbal, semantic, and problem solving skills they
had practiced for the five years before.  Assessment was both {\it
qualitative\/} -- e.g., asking 8th grade students to relate in their
own words why it is `that if you have two fractions with the same
numerator, the one with the smaller denominator is the larger'; and
{\it quantitative\/} -- e.g., administration of standardized
arithmetic examinations to test and control groups in the 6th
grade. Is it time for a science counterpart of the
Benezet/Berman Manchester experiment of the 1930's?
\endabstract

\parskip=\medskipamount \listleftindent=\a \listrightindent=\a

\section{Rote learning: Opium for the mind}

Traditionally taught science and mathematics teach little
except obedience.  Here are examples for the skeptical.
This gem is discussed by Alan Schoenfeld (1987):
\beginquote
  One of the problems on the NAEP [National Assessment of Educational
  Progress] secondary mathematics exam, which was
  administered to a stratified sample of 45,000 students nationwide, was
  the following: An army bus holds 36 soldiers.  If 1128 soldiers are
  being bused to their training site, how many buses are needed?

  Seventy percent of the students who took the exam set up the correct
  long division and performed it correctly.  However, the following are
  the answers those students gave to the question of `how many buses are
  needed?': 29\%\ said...31 remainder 12; 18\%\ said...31; 23\%\
  said...32, which is correct.  (30\%\ did not do the computation
  correctly).

  It's frightening enough that fewer than one-fourth of the students got
  the right answer.  More frightening is that {\it almost one out of three
  students said that the number of buses needed is `31 remainder
  12'}. [emphasis added]
\endquote

Mathematics has no meaning for most students; it is a sequence of
mysterious steps, which the clever or obedient quickly master to score
high on examinations.  More disturbing than mindless mathematics are
examples showing that students are not bothered by nonsense.  An
example from Schoenfeld (1991) shows the trouble:
\beginquote
In his dissertation research,
Kurt Reusser \dots (Reusser 1988)\dots
asked 97 first and second grade students the 
following question:  `There are 26 sheep and 10 goats on a ship.  How 
old is the captain?'  Seventy-six of the 97 students `solved' the 
problem, providing a numerical answer by adding 26 and 10.
\endquote
\noindent 
School aggravates the problem.
Into sets of problems worked by school children,
Radatz (cited in Schoenfeld (1989))
inserted non-problems such as the following (in its entirety!):
\beginquote
Mr.~Lorenz
and 3 colleagues started at Bielefeld at 9 a.m. and drove the 360 km
to Frankfurt, with a rest stop of 30 minutes.
\endquote
\noindent  The percentage of
students who answer such non-problems {\it increases consistently from
kindergarten through 6th grade}.
The more time students spend in rote 
learning, the more brain-dead they become.

The problem is international.
One of the authors (SM) gave this problem to a sample of physics
majors at the University of Cambridge:
\beginquote
Two people are on opposite
sides of a rotating merry-go-round.  One throws a ball to the other.
In which frame of reference is the path of the ball straight when
viewed from above? Choices: (a) the earth, (b) the merry-go-round, (c)
both, or (d) neither.
\endquote
\noindent Only 58\%\ of physics majors answered correctly; of whom
only 40\%\
were sure of their answer.  The success rate did not increase with
more years of studying physics.  It seems that
even very talented physics students depend on rote learning.\note{More
details of the questions and answers are on the web at
\url{http://www.inference.phy.cam.ac.uk/sanjoy}.}

\section{Benezet's (1935a, 1935b, 1936) curriculum}

Etta Berman, a teacher who worked under Benezet as a K-12 teacher in
Manchester, and researched the Benezet method for her master's thesis
(Berman 1935), reports an interview in which Benezet commented with
characteristic pungency:
\beginquote
We have been chloroforming children's reasoning powers.  We have been
drilling them in formulae and tables to the detriment of their
reasoning ability.  (Berman 1935, p.~45)
\endquote
\noindent Berman summarized Benezet's views as:
\beginquote
\dots greater intellectual powers can be secured by warding off material
which makes for mental stunting and substituting in its place content
in which the children find enjoyment, as well as things common to
their understanding, experience, and environment.

\dots formal arithmetic drilled before the child's reasoning powers
are developed is one of the underlying causes for stunted reasoning
powers.  (Berman 1935, p.~46)
\endquote
\noindent In two sentences Benezet described his method:
\beginquote
Teachers were told to soft-pedal the mechanical arithmetical drills
and to concentrate on reasoning, estimation, and on self-expression.
They were asked to give the children a great number of books to read,
and to encourage oral English in telling the story of these books.
(Berman 1935, p.~46)
\endquote

\medbreak Students learnt mathematics 
{\it in context}.\note{\noindent A warning to
those thinking of trying the method or of criticizing it for avoiding
all arithmetic: `Some teachers, new in the experiment, have steered
clear of all arithmetic but Mr. Benezet's idea is really to teach any
arithmetic as the need comes up but not on the plane of formal drill.'
(Berman 1935, p.~48)} First-graders meet small numbers:\beginquote
This instruction is not concentrated into any particular period or
time but comes in incidentally in connection with assignments of the
reading lesson or with reference to certain pages of the
text.\endquote
\noindent Second graders make friends
with larger numbers (their readers are longer):
\beginquote
If any book used in this grade 
contains an index, the children are taught what it means and how to 
find the pages referred to. Children will naturally pick up counting 
in the course of games which they play. 
\endquote
The principle of learning in context continues through the entire
curriculum.  A short summary of each year's curriculum is in 
\ref{table:Benezet}; detailed descriptions are in (Benezet 1935b).

\begintable{table:Benezet}{Benezet's curriculum.}
\everycr={\noalign{\smallskip}}%
\halign{#\hfil\quad&\vtop{\noindent\hsize=5truein #}\cr
\1
\it Grade&\hfil\it Mathematics ideas\cr
\2
1&Numbers $<100$; comparison: more, less, higher, lower\cr
2&Telling time (hours and half-hours) page numbers; using an index.\cr
3&Bigger numbers: license plates, house numbers.\cr
4&Inch, foot, yard.  Estimating lengths.  Square inch, square foot.\cr
5B&Counting by 5's, 10's, 2's, 4's, and 3's (mentally), leading to those
multiplication tables.  Estimation games; always writing estimate
before checking.  Fractions by pictures.\cr
5A&Multiplication table.\cr
6&Formal arithmetic, but {\it estimate first\/} then check.\cr
7&Lots of mental arithmetic without reference to paper or
blackboard.\cr
8B&More mental arithmetic.\cr
8A&Reasons for processes.  Explaining how to attack problems.\cr
\3}\endtable

\section{Assessment}

Benezet combined qualitative and quantitative assessment, the better to
convince people who accept one or other kind of evidence.  

As a qualitative assessment, he asked students questions
that appear 
ridiculously easy:
\beginquote
The distance from Boston to Portland [Maine] by water is 120 miles. 
Three steamers leave Boston, simultaneously, for Portland. One makes 
the trip in 10 hours, one in 12, and one in 15. How long will it be 
before all 3 reach Portland? (Benezet 1936)
\endquote
\noindent
In the regularly taught 9th grade, 6 out of the 29 students (!) got it
right.  In the experimental 2nd grade, all students got it right.
Benezet does not say what answer the regularly taught
9th-grade students gave, but the
example of sheep and goats suggests that it was probably 37.

Benezet also asked students to explain their thinking:
\beginquote
I was trying to get the children to tell me, in their own words, that
if you have two fractions with the same numerator, the one with the
smaller denominator is the larger.
\endquote
From the regularly taught 8th grade, he got: `The smaller number in
fractions is always the larger', and `The denominator that is smallest
is the largest.'

Etta Berman, in her Masters thesis (1935) on the Benezet experiment,
used numerous quantitative assessments: 
\unorderedlist
\li verbal: antonyms, synonyms, alphabet, analogies, anagrams
\li mathematical: word problems,
single- and multidigit addition and multiplication; subtraction, addition,
multiplication, and division of fractions; long division
\li everyday knowledge: `common sense'; typical prices (e.g. of
gasoline and coal),
interest rates, and discounts
\endunorderedlist
\noindent In formal
arithmetic, the experimental students caught up to the regular ones in
only 4 months during the 6th grade.  As Berman concluded with academic
understatement: `The results of this study cast doubt upon whether we
are justified in devoting five years to the drilling of formal
arithmetic' (Berman 1935, p.~40).  The experimental students instead
got years of extra practice in reading, writing, and thinking:
\beginquote
In one [traditionally taught]
fourth grade I could not find a single child who would admit
that he had committed the sin of reading. I did not have a single
volunteer, and when I tried to draft them, the children stood up,
shook their heads, and sat down again. In the four experimental fourth
grades the children fairly fought for a chance to tell me what they
had been reading. The hour closed, in each case, with a dozen hands
waving in the air and little faces crestfallen, because we had not
gotten around to hear what they had to tell.
(Benezet 1935a)
\endquote

\section{Teaching for transfer}

The examples of the army buses and of the steamers going to Portland
show that traditionally taught students cannot use their knowledge
except in rote, classroom problems.  In traditional mathematics
teaching, the applications are similar to one another, so the ideas
common to all the applications include much besides the essentials.
Thus the student cannot easily abstract the essential, transferable
ideas (\ref{fig:transfer}).  For an idea to transfer, students must
apply it in widely differing contexts.  Bransford, Brown, and Cocking
(1999, ch.~3) give a valuable discussion of transfer; see especially
the references on how `contrasting cases' enhance transfer (p.~48).
By having students read stories, poems, and adventures, by having them
read geography and history, and by having them invent their own
stories, Benezet introduced students to a vast diversity of
mathematical applications.  In this rich environment, students learnt
mathematical skills useful well beyond the classroom.

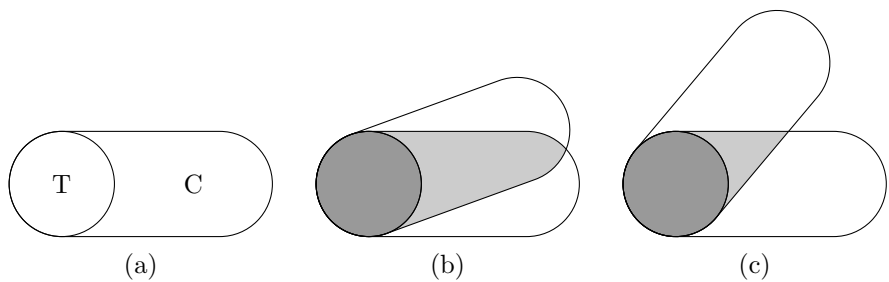
\mpsfig{fig:transfer}{fig.1}{Why `knowledge'
from traditional mathematics teaching does not transfer.  (a) Ideas
used in an application, showing contingent, contextual ideas (C) and
essential ideas (T), the ones of general use that you would like students
to transfer automatically to other contexts.  (b) Traditional
course with two applications.  The contexts are similar, so the
intersection of the two applications (shaded area) includes much more
than the essentials of the transferable ideas (darker shaded area).
(c) Teaching for transfer.  Here the contexts are diverse, so the
intersection is mostly the essentials of the ideas.}

\section{Mathematical proficiency}

In a recent National Research Council report,
{\it Adding It Up: Helping Children Learn Mathematics},
Kilpatrick et al. (2001) advise revamping P--8 (ages 3--12)
mathematics education
so that all students attain `mathematical proficiency': 
\unorderedlist
\def\i#1:{\li {\it #1:\/}}
\i conceptual understanding:
comprehension of mathematical concepts, operations, and relations 

\i procedural fluency:
skill in carrying out procedures flexibly, accurately, efficiently,
and appropriately

\i strategic competence:
ability to formulate, represent, and solve mathematical problems

\i adaptive reasoning:
capacity for logical thought, reflection, explanation, and
justification

\i productive disposition:
habitual inclination to see mathematics as sensible, useful, and
worthwhile, coupled with a belief in diligence and one's
own efficacy
\endunorderedlist
\noindent
According to committee chair Jeremy Kilpatrick (as quoted by Mervis
2001): `We want to move past the \dots [`Math Wars']\dots debate over
skills versus understanding.  It's not one or the other.  The point is
that both are needed, and more, to learn and understand mathematics.'

Berman's (1935) study suggests that Benezet's 8th-grade students
possessed far greater mathematical proficiency in the above sense than
did regularly taught students.  Perhaps the Math Wars can end in a
Treaty of Benezet, in the spirit of {\it Adding It Up}, and acceptable
to both sides, e.g., to the `Mathematically Correct'
\url{http://mathematicallycorrect.com} and the `Mathematically Sane'
\url{http://mathematicallysane.com}.

The requirements for mathematical proficiency given by Kilpatrick et
al. (2001) are consistent with Benezet's curriculum (Table 1); they
complement the Benezet/Swartz physics topics (Tables 2 and 3); and
they mesh with the features of science literacy listed by Arons
(1990).

\section{Physics}

A physics counterpart of the Benezet experiment could include a huge
variety of ideas.  In \ref{table:physics} we sketch a possible curriculum.
\begintable{table:physics}{A possible physics curriculum.}
\halign{#\hfil\quad&\vtop{\noindent\hsize=5truein #}\cr
\1
\it Grade&\hfil \it Physics and mathematics ideas\cr
\2
1--8&
Benezet's mathematics program as a basis.  Add {\it physical
quantities\/} to it: angles, volumes, weight (mass), force
(estimating only), density as students learn division, energy, power.
All quantities are related to everyday experience: density of rocks,
volume of houses, power required in climbing stairs or cycling, power
in the falling water at Niagara Falls.

Proportional reasoning and scaling: `If you double the side of a
cube, what happens to the volume?'  Use scaling throughout, starting
in grade 4, with the introduction of square measure, and emphasize it
especially in grade 8.\cr
9-11&Gravitation, motion of planets by hand simulation
to develop a tick-by-tick model of how Newton's second law works.
Dimensions, units, dimensional analysis
to guess formulae.  Springs, waves, sound, music, pressure.  Matter is
made of atoms.  This order would fit with a 
physics-first curriculum (Livanis 2001)
as suggested by Leon Lederman (1999).\cr
12&Begin exact calculations, including conservation laws.\cr
\3}\endtable

Qualitative assessment could include estimation problems:
\unorderedlist
\li How many miles must I run to burn off the calories from a
candy bar?
\li How high can animals jump as a function of size?
\li Why are hummingbirds so tiny?
\li Estimate for your country the fractional change in the
average miles per gallon (or km/liter) of gasoline used by
automobiles if the speed limit is changed from $55\,\rm mph$
(or $88\,\rm km/hr$) to $75\,\rm mph$ (or $120\,\rm km/hr$).
\endunorderedlist

\noindent High-school students could 
try to find flaws in perpetual-motion
machines.  Here are two such machines:
\numberedlist
\interitemskipamount=2pt
\li {\it Bouncing}.
A ball bounces elastically off a wall (in one dimension): It strikes
and rebounds with velocity $V$ and energy is conserved.  In a
reference frame that moves with the initial velocity of the ball, the
ball before the bounce has velocity 0; and after the bounce, velocity
$2V$.  In this frame, energy is not conserved.  Are you worried?

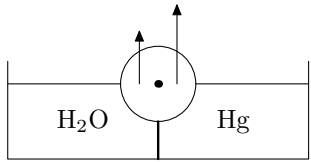
\li {\it Buoyancy}.
$$\rawmps{fig.2}$$
The figure shows a cross-section of a long trough, showing a cylinder
free to rotate about the spindle at its center, but not free to move
off axis.  The mercury and water are kept from mixing by an
impermeable membrane.  The upward arrows show the buoyant forces on
each half of the cylinder, with the mercury's contribution greater
than the water's by a factor of 13 (the density ratio).  So the
spindle should rotate and, by symmetry, keep rotating.
Perpetual motion?
\endnumberedlist
\bigbreak

After completing \ref{table:physics}, we stumbled across an editorial
by Cliff Swartz (1993). Swartz suggests topics (Table 3) for K--12
science education that seem compatible with the physics and
mathematics ideas of \ref{table:physics}.  His description of 
how to chose the important topics reminds us of Benezet:
\beginquote
Here are some samples of my proposed standards for physics
topics\dots The theme is that\dots all physics standards K--12 should
emphasize the quantitative\dots I mean dealing with the size of
things, relating numbers to measurable and measured magnitudes,
necessarily therefore paying attention to significant figures and
knowing how to do order-of-magnitude, zeroing-in
calculations\dots my elementary-school science standards involve
only topics and experiences that are literally tangible. The change to
abstraction should be proceed only during junior high, in general to
be introduced only after concrete examples\dots The National
Committee, NSTA, AAAS, Albert Shanker, and everyone else is making
lists of standards. Physics teachers should produce their own list
before some power from on high congeals a new federal dogma in our
name.
\endquote

\begintable{table:Swartz-topics}{Possible topics for K--12 suggested by
Swartz (1993).}
\halign{#\hfil\quad&\vtop{\noindent\hsize=5truein #}\cr
\1
\it Grade&\hfil \it Physics and mathematics topics\cr
\2
1--6&Use standard measuring tools.

Students measure their foot lengths, then organize and
interpret a class distribution graph.

Students time their pulses, plot a distribution of class results, and
make comparisons with results of before and after physical
exercise.

Select the needed apparatus and make all the measurements,
calculations, and graphs to determine who runs faster in their class,
tall kids or short kids.\cr

7--9&Use standard measuring tools.

Use wire,
battery, and bulb with the right tools and connectors to make the bulb
light. 

With a convex lens as a magnifier, produce both real and virtual
images. 

Measure the volume and mass of an object and calculate its
density. 

Measure work input and output of a simple machine. 

Use echoes
to measure the speed of sound.\cr
10--12&Use standard measuring tools.

Given the mass of a pollutant in a quantity of water, calculate the
degree of pollution in parts per million.

Organize the history of the universe on a power-of-ten map.

Characterize the electromagnetic spectrum in terms of wavelengths,
frequencies, and photon energies, doing necessary calculations and
examples to illustrate each regime.

Use Archimedes' principle to explain how a boat floats.\cr
\3}\endtable

What is the relationship of the Benezet/Swartz material in
\refs{table:physics} and \refn{table:Swartz-topics}
to the physical-science portion of the National Science Education
Standards (National Research Council 1996a) shown in
\ref{table:NSS}?
The Benezetian items in 
\refs{table:physics} and \refn{table:Swartz-topics}
are concrete examples of general categories in \ref{table:NSS}. 
In our view,
\refs{table:Benezet} and \refn{table:physics} may give teachers a
clearer idea of possible nuts and bolts for a worthwhile K--12 science
and mathematics education.

\begintable{table:NSS}{Physical science portion of the National
Science Education
Standards (National Research Council 1996a, Table 6.2, p.~106).}
\halign{#\hfil\quad&\vtop{\noindent\hsize=5truein #}\cr
\1
\it Grade&\hfil \it Physical science standards\cr
\2
K--4&Properties of objects and
materials.

Position and motion of objects.

Light, heat, electricity, and
magnetism.\cr
5--8&Properties and changes of properties in matter

Motions and forces

Transfer of energy\cr
9--12&Structure of atoms

Structure and properties of matter

Chemical reactions

Motions and forces

Conservation of energy and increase in disorder

Interactions of energy and matter\cr
\3}\endtable

\medbreak
After considering Benezet's articles, the late Arnold Arons (2000)
wrote to us:
\beginquote
I have looked at the Benezet papers, and I find the story congenial. 
The importance of cultivating the use of English cannot be 
exaggerated. I have been pointing to this myself since the '50's, and 
am delighted to find such explicit agreement. I can only point out 
that my own advocacy has had no more lasting effect than Benezet's. 
You will find some of my views of this aspect in my 1959 paper
\dots(Arons 1959)\dots on the Amherst course. 
\dots{\it
It is worth noting 
that what Benezet was doing 70 years ago could now be done even more 
effectively because of the existence of excellent science and social 
studies curricula\dots(Arons 1993, 1997, 1998)\dots that would 
provide rich opportunity for invoking and applying just the kind of 
thinking, reasoning, and interpretation that Benezet was advocating 
and for which he had to invent his own illustrations.} [emphasis added]
\endquote
\noindent
Consistent with Arons's emphasis, Appendices I and II provide
post-1930's resources and references, respectively, that may assist a
Benezetian overhaul of K--12 science education. Many of the listed 
items are taken from Hake (2000a), which has
even more resources and references.

Quantitative assessment could include various `Classroom Assessment
Techniques' (NISE 2001) and diagnostic tests of conceptual
understanding (Hake 2001b, NCSU 2001). Unfortunately, there are, as far
as we know, few research-based diagnostic tests for the measurement of
science understanding in grades K--10.

\section{So what?}

Sixty years ago, in a small New England mill town, a creative
superintendent made mathematics meaningful for students.  Students
today still leave traditional mathematics and science classes with
almost no conceptual understanding of the subjects and little ability
to solve real-world problems. For too long the pioneering experiment
of Benezet has been ignored by the education community.

We hope the reader will forgive the length of following quotation,
which describes teaching that we all can admire.  Berman attended a
demonstration lesson in which Benezet taught one of the experimental
classrooms:
\beginquote
\dots the children were on their feet every chance they could
possibly get to tell all they knew.  They were free to think and,
indeed, made the most of every opportunity.  One could see mental
expansion in actual operation, and also observe satisfactions
register, when accredited with right answers or worthwhile opinions[,]
for every second brought something to challenge their interests, and
which they had to reach high to get.  The teachers present saw no
parasites but rather children who were taking a delight in independent
intelligent thinking.  These children were very much alive and their
was not a single instance of a child being self-conscious or wishing
to be over-looked.  These children were free and unhampered by formal
procedure and were finding themselves a part of the great world and
scheme of things in which their knowledge and opinions mingled with
those of older folks.  They were transmitting and acquiring knowledge,
and by means of directed guessing, when in doubt, were developing
judgment.  These children of a sixth grade were, indeed, laying the
foundation for worthwhile thinking and judgment which is so essential
to intelligent citizenship.  These active and thinking individuals not
only transmitted and received knowledge but also lived and felt the
things they talked about because they were being brought up on
Mr. Benezet's formula for {\it intellectual curiosity}. [emphasis
original] (Berman 1935, p.~50)
\endquote

\noindent So YES! It is time for a science counterpart of the
Benezet--Berman math experiment of the 1930's.  {\bf Why not give it a
try in your classroom?}

\section{Acknowledgements}

We are indebted to David Brown, Jack Lochhead, and the late Arnold
Arons for valuable comments on the work of Benezet.

\parskip=0pt \parindent=\a

\bigskip\bigskip

\lsection{sec:references}{References}

\begingroup \refdefs
Arons, A.B. (1959). Structure, methods, and objectives of the 
required freshman calculus-physics course at Amherst College.
\j{American Journal of Physics}{27\rm(9)}{658--666}.

Arons, A.B. (1990).  \t{A Guide to Introductory Physics Teaching}.
Wiley.  Reprinted with minor updates as \t{Teaching Introductory
Physics\/} (Wiley, 1997), which also contains \t{Homework and Test
Questions for Introductory Physics Teaching\/} (Wiley, 1994) along
with a new monograph \t{Introduction to Classical Conservation Laws}.
Both the 1990 and 1997 versions contain `Achieving wider scientific
literacy' (Chapter 12), with Arons's twelve `marks of sicnetific
literacy'.  These 12 marks are also online in Hake (2000b), pp. 5--7.

Arons, A.B. (1993). Uses of the past: reflections on United States
physics curriculum development, 1955 to
1990. \j{Interchange}{24\rm(1\&2)}{105--128}.

Arons, A.B. (1997). Improvement of physics teaching in the heyday of
the 1960's. In J. Wilson (ed.), \t{Conference on the introductory
physics course on the occasion of the retirement of Robert Resnick},
Wiley, pp. 13--20.

Arons, A.B. (1998). Research in physics education: The early years. 
In T.C. Koch and R.G. Fuller, eds.,
\t{PERC 1998: Physics Education Research Conference Proceedings 1998},
online at \url{http://physics.unl.edu/perc98}.

Arons, A.B. (2000).  Private communication to R.R. Hake (30 June).

Benezet, L.P. (1935a).  The Teaching of Arithmetic I: The Story
of an Experiment.  \j{Journal of the National Education
Association}{24}{241--244} (November).  Reprinted 
in the \j{Humanistic Mathematics Newsletter}{6}{2--14} (May 1991)
and on the web at the Benezet Centre 
\url{http://www.inference.phy.cam.ac.uk/sanjoy/benezet}.

Benezet, L.P. (1935b).  The Teaching of Arithmetic II: The Story
of an Experiment.  \j{Journal of the National Education
Association}{24}{301--303} (December).  Reprinted 
in the \j{Humanistic Mathematics Newsletter}{6}{2--14} (May 1991)
and on the web at the Benezet Centre
\url{http://www.inference.phy.cam.ac.uk/sanjoy/benezet}.

Benezet, L.P. (1936).  The Teaching of Arithmetic III: The Story
of an Experiment.  \j{Journal of the National Education
Association}{25}{7--8} (January).  Reprinted 
in the \j{Humanistic Mathematics Newsletter}{6}{2--14} (May 1991)
and on the web at the Benezet Centre
\url{http://www.inference.phy.cam.ac.uk/sanjoy/benezet}.

Berman, Etta (1935).  \t{The Result of Deferring Systematic Teaching
of Arithmetic to Grade Six as Disclosed by the Deferred Formal
Arithmetic Plan at Manchester, New Hampshire}.  Masters Thesis, Boston
University, USA.

Bransford, J.D., A.L. Brown, and R.R. Cocking (eds.) (1999). \t{How
People Learn: Brain, Mind, Experience, and School}. National Academy
Press.  Online at \url{http://www.nap.edu/catalog/6160.html}.

Hake, R.R. (2000a). Is it finally time to implement curriculum S? 
\j{AAPT Announcer}{30\rm(4)}{103}.  Online at
\url{http://www.physics.indiana.edu/~hake}.

Hake, R.R. (2000b). The general population's ignorancce of
science-related societal issues: A challenge for the university.
\j{AAPT Announcer}{30\rm(2)}{105}.  Online at
\url{http://physics.indiana.edu/~hake}.

Hake, R.R. (2001b). Suggestions for administering and reporting
pre/post diagnostic tests. Unpublished. Online at
\url{http://physics.indiana.edu/~hake}.

Kilpatrick, J., J. Swafford, and B. Findell, eds. 2001.  \t{Adding It
Up: Helping Children Learn Mathematics}. National Academy Press.
Prepublication copy online at
\url{http://www.nap.edu/catalog/9822.html}.

Lederman, L.M. (1999). A science way of thinking.  {\it Education
Week\/} 16 June, \url{http://www.edweek.org/ew/1999/40leder.h18}.

Livanis, O.  (2000). Physics First Home Page;
  \url{http://members.aol.com/physicsfirst/}

Mervis, J. (2001).  Math education: Academy report aims to quiet
debate.  \j{Science}{291}{808} (2 February). Reports on the National
Research Council report of Kilpatrick et al. (2001).

National Research Council (1996a). \t{National Science Education
Standards}. National Academy Press.
\url{http://books.nap.edu/catalog/4962.html}.

NISE [National Institute for Science Education] (2001).
\url{http://www.wcer.wisc.edu/NISE}: Field-tested Learning Assessment 
Guide (FLAG): For Science, Math, and Engineering Instructors.
\url{http://www.wcer.wisc.edu/nise/cl1/flag/flaghome.asp}, especially
Introduction to CAT's [Classroom Assessment Techniques] at
\url{http://www.wcer.wisc.edu/nise/cl1/flag/cat/catframe.asp}.

NCSU [North Carolina State University] (2001).  Physics Education R
\&\ D Group, Assessment Instrument Information Page, online at
\url{http://www.ncsu.edu/per/TestInfo.html}.

Radatz, H. (1983).  Untersuchungen zum L\"osen eingekleideter
Aufgaben.  [Studies of solving mathematical tasks] \j{Journal f\"ur
Mathematikdidaktik}{3}{205--217}.

Reusser, Kurt (1988).  Problem Solving
beyond the Logic of Things: Contextual Effects on Understanding and
Solving Word Problems.  \j{Instructional Science}{17}{309--38}.

Schoenfeld, Alan (1987).  What's all the fuss about metacognition? In
\t{Cognitive Science and Mathematics Education}, Alan
Schoenfeld, ed. (Hillsdale, NJ: Lawrence Erlbaum).

Schoenfeld, Alan H. (1989).  Teaching mathematical thinking and
problem solving.  In Lauren B. Resnick and Leonard B. Klopfer, eds.,
\t{Toward the Thinking Curriculum: Current Cognitive Research},
Association for Supervision and Curriculum Development [ASCD],
Alexandria, VA.

Schoenfeld, Alan H. (1991). On Mathematics as sense-making: An
informal attack on the unfortunate divorce of formal and informal
mathematics.  In J.F. Voss, D.N.
Perkins, and J.W. Segal, eds.,
\t{Informal Reasoning and Education}, Erlbaum, Hillsdale, NJ,
pp.~311--344.

Swartz, C.E. (1993). Editorial: Standard reaction. \pt{31}{334--335}.

\endgroup

\section{Appendix I}

{\bf Resources for a Benezetian overhaul of K--12 science education.}

\noindent (A slash '/' occurring outside a URL $\langle\ldots\rangle$ means
`click on the following text'.)

\begingroup
\refdefs

AAHE's CASTL (2001). Carnegie Academy for the Scholarship of 
Teaching and Learning, Programs for K--12; online at 
\url{http://www.carnegiefoundation.org/CASTL/k-12/index.htm}.

AAPT's Physical Science Resource Center \url{http://www.psrc-online.org/}.

Active Learning Problem Sets (ALPS)\hb
\url{http://www.physics.ohio-state.edu/~physedu/people/vanheu/index.html}.

Active Physics \url{http://www.psrc-online.org/} /Curriculum/High
School/Comprehensive Curricula/; also 
\url{http://www.its-about-time.com/htmls/index3.html}.

ActivPhysics
\url{http://www.physics.ohio-state.edu/~physedu/people/vanheu/index.html}.

Alive Education.net \url{http://www.alincom.com/educ/sci.htm} `Excellence 
in Internet Education'.

The Coalition for Education in the Life Sciences [CELS], a 
`national coalition of professional
societies in the biological sciences that have joined together in an 
effort to improve undergraduate education
in the life sciences' at \url{http://www.wisc.edu/cels/}.

Comprehensive Conceptual Curriculum for Physics (C3P)
\url{http://www.udallas.edu/physics/}.

Constructing Ideas in Physical Science \url{http://cipsproject.sdsu.edu/}.

Constructing Physics Understanding (CPU)
\url{http://cpuproject.sdsu.edu/CPU}.  See also\hb
\url{http://learningteam.org/} /CPU.

Cooperative Group Problem Solving
\url{http://www.physics.umn.edu/groups/physed/}.

Dewey web sites:
  (a) by Craig Cunningham \url{http://cuip.uchicago.edu/~cac/dewey.html}.
(b) Center for Dewey Studies  \url{http://www.siu.edu/~deweyctr/}
(c) Hoover's Teacher Ed Pages -- Links to the World of John Dewey
     \url{http://www.cisnet.com/teacher-ed/dewey.html}

Educational Development Center \url{http://www.edc.org/}.  `Founded in 
1958 when a group of scientists at the Massachusetts Institute of 
Technology joined forces with teachers and technical specialists to 
develop a new high school physics curriculum, PSSC Physics.'

Eison, J.A. et al. Bibliography of Active Learning in Science.  At\hb
\url{http://www.cte.usf.edu/resources/res_def.html}.

Event-Based Science: `A new way to teach science at the middle school level'.
\url{http://www.mcps.k12.md.us/departments/eventscience/}.

Exemplary \&\ Promising Science Programs.  US Deptartment of Education.
\url{http://www.ed.gov/offices/OERI/ORAD/KAD/expert_panel/newscience_progs.html}.

Experiment Problems
\url{http://www.physics.ohio-state.edu/~physedu/people/vanheu/index.html}.

The Exploritorium Institute for Inquiry
\url{http://www.exploratorium.edu/IFI/index.html}.

Exploritorium Inquiry Bibliography: Professional Development Design
Seminar Guide to Articles and Readings
\url{http://www.exploratorium.edu/IFI/resources/biblio.html}.

Harvard-Smithsonian Center for Astrophysics -- Science Education
Department
\url{http://cfa-www.harvard.edu/cfa/sed/projects.html}.

Hands On Physics \url{http://hop.concord.org/}.

Insights, Center for Science Education \url{http://www.edc.org/CSE/}, 
developers of \t{Insights: An Inquiry-Based Elementary School Science 
Curriculum\/} (K--6)\hb \url{http://www.edc.org/CSE/imd/insights3.html} and 
\t{Insights: An Inquiry-Based, Middle School Science Curriculum\/}
\url{http://www.edc.org/CSE/imd/insights2.html}.

Lederman Science Center at Fermilab
\url{http://www-ed.fnal.gov/ed_lsc.html}.

Logo Foundation
\url{http://lcs.www.media.mit.edu/groups/logo-foundation/index.html}.
According to Harold Abelson, `Logo is the name for a philosophy of 
education and a continually evolving family of programming languages 
that aid in its realization.'

Mechanical Universe \url{http://www.projmath.caltech.edu/mu.htm}.

Minds On Physics 
\url{http://umperg.physics.umass.edu/projects/MindsOnPhysics/default}.

Modeling Instruction Program \url{http://modeling.la.asu.edu/modeling.html}.

NSF Review of Instructional Materials for Middle School Science
\hb\url{http://www.nsf.gov/pubs/1997/nsf9754/nsf9754.htm}.

Paideia Program \url{http://www.paideia.org/}.

Peer Instruction and Concept Tests \url{http://galileo.harvard.edu/}.

Physics InfoMall \url{http://learningteam.org/}.

Physics by Inquiry \url{http://www.phys.washington.edu/groups/peg/pbi.html}

Physics Instructional Resource Center \url{http://pira.nu/}.

Physics Resources and Instructional Strategies for Motivating Students (PRISMS)
\url{http://www.prisms.uni.edu/}.

Physics Teachers Resource Agents (PTRA)
\url{http://www.aapt.org/programs/ptra/ptra.html}.

Powerful Ideas in Physical Science \url{http://www.psrc-online.org/}
/Curriculum/College--University\1/Pre-service Teacher Education.

SEPUP Modular Materials
\url{http://www.lhs.berkeley.edu/SEPUP/general.html}.  For a discussion 
see Wilson \&\ Davis (1994), pp.~205--210.

Science Activities Manual: K--8
\url{http://www.utm.edu/departments/ed/cece/SAMK8.shtml}. 

Science Helper K--8 \url{http://learningteam.org/}:  
\beginquote
Funded by the 
Carnegie Corporation of New York, and created by a team of leading 
teachers, specialists, and administrators, Science Helper K--8 CD-ROM 
offers 919 lesson plans which are the culmination of 15 years of 
development, field-testing and refinement.  The lesson plans 
represent seven of the most effective and influential science 
curricula ever written -- COPES, ESS, ESSP, MINNEMAST, SAPA, SCIS, and 
USMES. You can easily locate the lesson you're looking for by grade 
level, subject, process skill, keyword or content... then print it 
out. All in all, Science Helper is the single most effective, most 
comprehensive way to expand your K--8 science program immediately.
\endquote

Socratic Dialogue Inducing (SDI) Labs
\url{http://www.physics.indiana.edu/~sdi}.

Tools for Scientific Thinking \url{http://www.vernier.com/cmat/tst.html}.

Tutorials in Physics \url{http://www.phys.washington.edu/groups/peg/tut.html}.

Workshop Physics \url{http://physics.dickinson.edu/} /Workshop Physics.

Workshop Science \url{http://physics.dickinson.edu/} /Workshop Science Project.

\endgroup  

\section{Appendix II}

{\bf References for a Benezetian overhaul of K--12 science
education}\note{A * preceding a reference indicates that
the reference is also given in \ref{sec:references}.}

\begingroup
\refdefs
\def\*{\noindent\llap{*}}

AAAS (1989) [American Academy of Arts and Sciences].  \t{Science for All
Americans: A Project 2061 Report on the Literacy Goals in Science,
Mathematics, and Technology}.  American Association for the Advancement
of Science.

AAAS (1993).  \t{Benchmarks for Science Literacy: Project 2061}.  American
Association for the Advancement of Science.

AAAS (1998). {\it Daedalus\/} {\bf 127\rm (4)}. For a description see 
\url{http://daedalus.amacad.org/inprint.html}. Contains essays by 
researchers in education and by historians of more rapidly developing 
institutions such as power systems, communications, health care, and 
agriculture. Sets out to answer a challenge posed by Kenneth Wilson: 
\beginquote
If other major American `systems' have so effectively demonstrated 
the ability to change, why has the educational `system' been so 
singularly resistant to change? What might the lessons learned from 
other systems' efforts to adapt and evolve, have to teach us about 
bringing about change -- successful change -- in America's schools?
\endquote
\hskip\a See also \url{http://www.physics.ohio-state.edu/~redesign/}.

Adey, P.S. (1999). The Science of Thinking and Science for Thinking: 
A Description of Cognitive Acceleration through Science Education 
(CASE). UNESCO, International Bureau of Education, Switzerland.
Online at\hb
\url{http://www.ibe.unesco.org/International/Databanks/Innodata/inograph.htm}.

Adler, M.J. (1977). \t{Reforming Education: The Opening of the American
Mind}.  Macmillan.

Allen, D.E. (1997).  Bringing Problem-Based Learning to the 
Introductory Biology Classroom.  In A.P. McNeal and C. D'Avanzo,
eds., \t{Student-Active Science: Models of 
Innovation in College Science Teaching}, Saunders.  Online at\hb
\url{http://www.saunderscollege.com/lifesci/studact/chapters/ch15.html}.

Allen, D.E. and B. Duch (1998). \t{Thinking Towards Solutions:
Problem-Based Learning Activities for General Biology}.  Saunders.

Allen, D.E., B.J. Duch, and S. E. Groh (1996). The power of 
problem-based learning in teaching introductory science courses. 
\j{New 
Directions for Teaching and Learning}{68}{43--52}.

American Mathematical Society Staff, American Institute of Physics
Staff, American Chemical Society (2001). \t{The Best of Wonder Science:
Elementary Science Activities}, Volume 2.  Wadsworth Publishing.
Recommended by Hubisz et al. (2001).

Anderson, L.W. and D.R. Krathwohl (2000). \t{Taxonomy for Learning,
Teaching, and Assessing: A Revision of Bloom's Taxonomy of Educational
Objectives}.  Longman.

Anderson, J.R. (1999).  \t{Cognitive Psychology and Its Implications}.
5th ed.  Especially the chapter on `Cognitive Development'.
W.H. Freeman.

Ansbacher, T. (2000). An interview with John Dewey on science
education. \pt{38\rm(4)}{224--227}.  A thoughtful and well-researched
treatment showing the consonance of Dewey's educational ideas -- as
quoted straight from Dewey's own writings, not from the views of
sometimes confused Dewey interpreters -- with the thinking of most
current science-education researchers, including your friendly
authors.

Apelman, M. (1993).  Co-creating primary curriculum: Boulder Valley
schools.  In E. Jones, ed., \t{Growing Teachers: Partnerships in Staff
Development\/} (National Association for the Education of Young
Children).  Online at
\url{http://www.clas.uiuc.edu/fulltext/cl01141/ch6.html}.

Apelman, M., D. Hawkins, and P. Morrison
(1985).  \t{Critical Barriers Phenomenon in
Elementary Science}.  University of North Dakota, Center for Teaching
and Learning.  See also Hawkins (1974c).

\*Arons, A.B. (1959). Structure, methods, and objectives of the 
required freshman calculus-physics course at Amherst College.
\j{American Journal of Physics}{27\rm(9)}{658--666}.

Arons, A.B. (1960). The New high school physics course: A report on 
PSSC.  {\it Physics Today}, June, pp.~20--25.

Arons, A.B. (1977).  \t{The Various Language: An Inquiry Approach to the
Physical Sciences}.  Oxford. With teachers' guide.

Arons, A.B. (1990). \t{A Guide To Introductory Physics Teaching},
Wiley.  Reprinted with minor updates in \t{Teaching Introductory
Physics\/} (Wiley, 1997), which also contains \t{Homework and Test
Questions for Introductory Physics Teaching} (Wiley, 1994) along with
a new monograph \t{Introduction to Classical Conservation Laws}.

Arons, A.B. (1991).  Guiding insight and inquiry in the introductory
physics lab. \j{Physics Teacher}{31}{278--282} (May).

\*Arons, A.B. (1993). Uses of the past: reflections on United States
physics curriculum development, 1955 to
1990. \j{Interchange}{24\rm(1\&2)}{105--128}.

\*Arons, A.B. (1997). Improvement of physics teaching in the heyday of
the 1960's. In J. Wilson (ed.), \t{Conference on the Introductory
Physics Course on the Occasion of the Retirement of Robert Resnick},
Wiley, pp. 13--20.

\*Arons, A.B. (1998). Research in physics education: The early years. 
In T.C. Koch and R.G. Fuller, eds.,
\t{PERC 1998: Physics Education Research Conference Proceedings 1998},
online at \url{http://physics.unl.edu/perc98}.

Arons, A.B. (1999). Development of energy concepts in introductory
physics courses.  \j{American Journal of Physics}{67\rm(12)}{1063}.

Arons, A.B. (1999). Closet foot-pounders? \j{American
Journal of Physics}{67\rm(6)}{468}.

Arons, A.B. and R. Karplus (1976).  Implications of accumulating data on
levels of intellectual development. \j{American Journal of
Physics}{44\rm(4)}{396}.

Asimov, I. and R.A. Galant (1973). \t{Ginn Science Program}. Ginn \&\
Company. For elementary schools -- many volumes, out of print. 
Recommended by Hubisz et al. (2001): `Has what is missing from most of
the books in this report.'

Aubrecht, G.J. (1991). Is there a connection between testing and
teaching? \j{Journal of College Science Teaching}{20}{152--157}.

Ausubel, D.P. (2000). \t{The Acquisition and Retention of Knowledge}.
Kluwer.

Ball, D.L. and H. Bass (2000).  Interweaving content and pedagogy in
teaching and learning to teach: Knowing and using mathematics.  In J.
Boaler, ed., \t{Multiple Perspectives on Mathematics Teaching and
Learning\/} (Ablex), pp. 83--104.

\*Benezet, L.P. (1935, 1936). The Teaching of Arithmetic I, II, III: 
The Story of an Experiment.  \j{Journal of the National Education 
Association}{24\rm(8)}{241--244} (1935); \nj{24\rm(9)}{301--303}
(1935); \nj{25\rm(1)}{7--8} (1936). 
The articles (a) were reprinted in the \j{Humanistic Mathematics 
Newsletter}{6}{2--14} (May 1991); (b) are
on the web along with other Benezetia at the 
Benezet Centre
\url{http://www.inference.phy.cam.ac.uk/sanjoy/benezet/}.

\*Berman, Etta (1935).  \t{The Result of Deferring Systematic Teaching
of Arithmetic to Grade Six as Disclosed by the Deferred Formal
Arithmetic Plan at Manchester, New Hampshire}.  Masters Thesis, Boston
University, USA.

Black, P. and D. Wiliam (1998). The black box: Raising standards through
classroom assessment. \t{Phi Delta Kappan}, October, online at
\url{http://www.pdkintl.org/kappan/kbla9810.htm}: 
\beginquote
Firm evidence shows 
that formative assessment is an essential component of classroom work 
and that its development can raise standards of achievement, Mr. 
Black and Mr. Wiliam point out. Indeed, they know of no other way of 
raising standards for which such a strong {\it prima facie\/} case can be 
made.
\endquote

Bloom, B.S. (1984). The 2-sigma problem: The search for methods of
group instruction as effective as one-to-one tutoring. \j{Educational
Researcher}{13\rm(6)}{4--16}.

Bohren, C.F. (1991). \t{What Light Through Yonder Window Breaks?: More
Experiments in Atmospheric Physics}.  John Wiley.

Bohren, C.F. (2001). \t{Clouds in a Glass of Beer: Simple Experiments
in Atmospheric Physics}. Dover.

\*Bransford, J.D., A.L. Brown, and R.R. Cocking (eds.) (1999). \t{How
People Learn: Brain, Mind, Experience, and School}. National Academy
Press.  Online at \url{http://www.nap.edu/catalog/6160.html}.

Brown, D. E. (1993). Refocusing core intuitions: A concretizing role
for analogy in conceptual change.  \j{Journal of Research in Science
Teaching}{30\rm(10)}{1273--1290}.

Brown, D.E. (2000). Merging dynamics: An integrating perspective on
learning, conceptual change, and teaching, and its implications for
educational research.  Paper presented at AERA 2000, New Orleans.
Online at \url{http://faculty.ed.uiuc.edu/debrown/merge/}.

Brown, J.S., A. Collins, and P. Duguid (1989). Situated cognition and
the culture of learning.  \j{Educational
Researcher}{18\rm(1)}{34--41}.  Online at\hb
\url{http://www.ilt.columbia.edu/ilt/papers/JohnBrown.html}.

Bruer, J.T. (1994). \t{Schools for Thought: A Science of Learning in
the Classroom}. MIT Press.

Bruer, J.T.  (1997).  Education and the brain: A bridge too far.
\j{Educational Researcher}{26\rm(8)}{4--16}.

Bruer, J.T. (1999). In search of\dots brain-based education. \t{Phi 
Delta Kappan}, May, p. 648.  Online at 
\url{http://www.pdkintl.org/kappan/kbru9905.htm}.

Bruer, J.T. (1999). \t{The Myth of the First Three Years}. Free
Press. See also
\url{http://www.jsmf.org/zarticles&pap/John/myth_of_the_first_three_years.htm}

Bunge, M. (1998).  \t{Philosophy of Science: From Explanation to
Justification}. Science \&\ Technology Series, Transaction Pub.

Bunge, M.  and N. Mahner, ed. (2001). \t{Scientific Realism: Selected
Essays of Mario Bunge}. Prometheus Books.

Butterworth, B. (1999). \t{What Counts: How Every Brain Is Hardwired
for Math}. Free Press, New York.  Published in England as \t{The
Mathematical Brain}, Macmillan, London, 1999.

Bybee, R.W. (1991).  Planet Earth in Crisis: How Should Science
Educators Respond?  \j{American Biology Teacher}{53\rm(3)}{146--153}.

Bybee, R.W. (1993). \t{Reforming Science Education: Social
Perspectives \&\ Personal Reflections}. Teachers College Press.

Chabay, R.W. and B.A. Sherwood (1999). Bringing atoms into first-year
physics.  \j{American Journal of Physics}{67\rm(12)}{1045}.

Chin, C. and D.E. Brown (2000). Learning in science: A comparison of
deep and surface approaches. \j{Journal of Research in Science
Teaching}{37}{109--138}.

Clement, J.M. (2000). Using physics to raise thinking skills.  \j{AAPT 
Announcer}{30\rm(4)}{81}.

Collins, A.  J.S. Brown, and S. Newman (1989). Cognitive
apprenticeship: Teaching students the craft of reading, writing, and
mathematics.  In L.B. Resnick, ed., \t{Knowing, Learning, and
Instruction: Essays in Honor of Robert Glaser}, Erlbaum, pp.~453--494.

Cooper, J., and P. Robinson (1998).  Small group instruction in science,
mathematics, engineering, and technology: A discipline status report
and teaching agenda for the future.  \t{Journal of College Science
Teaching}, May, pp.~383--388.

Csikszentmihalyi, M. (1990). Literacy and intrinsic motivation. 
\j{Daedalus}{119\rm(2)}{115--140}.
\beginquote
The chief impediments to learning are not cognitive. It is not that
students cannot learn; it is that they do not wish to. If educators
invested a fraction of the energy they now spend trying to transmit
information in trying to stimulate the students  enjoyment of
learning, we could achieve much better results.
\endquote

D'Avanzo, C.  and A.P. McNeal, eds. (1997). \t{Student-Active Science: 
Models of Innovation in College Science Teaching}.  Saunders. Saunders 
is advancing education reform: (a) `professionals' may 
order a free copy by calling +1$\,$800.782.4479; (b) the entire book will 
soon be available electronically as an Adobe Acrobat portable 
document file!
\url{http://www.saunderscollege.com/lifesci/studact/}.

Dewey, J. (1996). \t{Collected Works of John Dewey, 1882--1953\/}.
Southern Illinois University Press.  On CDROM; for information see
\url{http://www.siu.edu/~deweyctr/index.html}.

Dewey, J. (1997).  \t{Democracy and Education: An Introduction to the
Philosophy of Education}.  Simon \&\ Schuster.

Dewey, J. (1997). \t{Experience and Education}. Macmillan.

Dewey, J. and R.D. Archambault (ed.) (1983). 
\t{John Dewey on Education}. University of
Chicago Press.

Dewey, J. and Jo Ann Boydston (contributor) (1980). \t{The School and
Society}. Southern Illinois University Press.

Dewey, J., L. Hickman,
T.M. Alexander (eds.) (1998). \t{The Essential Dewey: Pragmatism,
Education, Democracy}. Indiana University Press.

Dewey, J. and P.W. Jackson (illustrator) (1991). \t{The School and
Society and the Child and the Curriculum}. A Centennial Publication;
University of Chicago Press.

diSessa, A.A. (1987). The third revolution in computers and 
education.  \j{Journal of Research in Science Teaching}{24}{343--367}.

diSessa, A.A. (2000). \t{Changing Minds: Computers, Learning, and
Literacy}.  MIT Press.

Dobey, D.C., R.J. Beichner, and S.L. Raimondi (1998). \t{Essentials of
elementary science}.  Allyn \&\ Bacon.  Recommended by Hubisz et
al. (2001).

Druckman, D. and R.A. Bjork, eds. (1994). \t{Learning, Remembering, and
Believing: Enhancing Human Performance}. National Academy Press. See
especially the Epilogue, `Institutional Impediments to Effective
Training'.

Duckworth, E., J. Easley, D. Hawkins, and A. Henriques, eds. (1990).
\t{Science Education: A Minds-on Approach for the Elementary Years}. 
Erlbaum.

Duckworth, E.  (1996).  \t{The Having of Wonderful Ideas and Other
Essays on Teaching and Learning}.  Teachers' College Press.

Elby, A. (1999). Another reason that physics students learn by rote. 
\t{Physics Education Research\/} supplement 1 to the 
\j{American Journal of Physics}{67\rm(7)}{S52--S57}.

Elmore, D.A. , S.E. Olsen, and P.M. Smith (1998).  \t{Reinventing
Schools: The Technology is Now}. National Academy Press.  Available
only on the web at\hb
\url{http://www.nap.edu/readingroom/books/techgap/pdf.html}.

Epstein, J. (1997/8). Cognitive development in an integrated
mathematics and science program.  \t{Journal of College Science
Teaching}, December 1997 and January 1998, pp.~194--201.

Ericsson, K.A., R.T. Krampe, and C. Tesch-Romer (1993). The role of 
deliberate practice in the acquisition of expert performance. 
\j{Psychological Review}{100\rm(3)}{363--406}.

Ericsson, K.A. and N. Charness (1994). Expert performance, its structure
and acquisition.  \j{American Psychologist}{49}{725--747}.

Ericsson, K.A. and J. Smith, eds. (1991). \t{Toward a General Theory of
Expertise: Prospects and Limits}.  Cambridge University Press.

Ericsson, K.A, ed. (1996).  \t{The Road to Excellence: The Acquisition
of Expert Performance in the Arts and Sciences, Sports, and Games}.
Erlbaum.

Fawcett, H.P. (1938). \t{The Nature of Proof: A Description and
Evaluation of Certain Procedures Used in a Senior High School to
Develop an Understanding of the Nature of Proof}.  Teachers' College
Press.  It was the 1938 \t{Yearbook\/} of the National Council of
Teachers of Mathematics and is now available from them as
H.P. Fawcett, {\it The Thirteenth Yearbook: The Nature of Proof},
1995.  For a discussion of Fawcett's brilliant work see Schoenfeld
(1991).

Fisher, K.M., J.H. Wandersee, and D. Moody (2000). \t{Mapping Biology
Knowledge}.  Kluwer.

Fiske, E.B. (1992). \t{Smart Schools, Smart Kids}. Touchstone.

Flener, F. (2001).  A geometry course that changed their lives: the
guinea pigs after 60 years.  Talk presented on the work of Fawcett
(1938) at the National Council of Teachers of Mathematics on 6 April
2001, Orlando, Florida. Online at
\url{http://groups.yahoo.com/group/math-learn/files/NatureOfProof.txt}.
Unfortunately, one must either be a subscriber to math-learn
\url{http://groups.yahoo.com/group/math-learn} or else sign in as a
subscriber to access the talk.

Ford, K.W. (1989).  Guest Comment: Is physics difficult?  
\j{American Journal of Physics}{57\rm(10)}{871--872}.  For
quotes from this insightful piece regarding physics in the early
grades see Hake (2000): \beginquote Physics is difficult in the same
way that all serious intellectual effort is difficult. Solid
understanding of English literature, or economics, or history, or
music, or biology or physics does not come without hard work. But we
typically act on the assumption (and argue to our principals and
deans) that ours is a discipline that only a few are capable of
comprehending. The priesthood syndrome that flows from this assumption
is, regrettably, seductive\dots If physics is not more difficult than
other disciplines, why does everyone think that it is? To answer
indirectly, let me turn again to English. Six-year-olds write English
and (to pick a skilled physicist writer) Jeremy Bernstein writes
English. What separates them? A long, gradual incline of increased
ability, understanding, and practice. Some few people, illiterates, do
not start up the hill. Most people climb some distance. A few climb as
far as Bernstein. {\it For physics, on the other hand, we have
fashioned a cliff. There is no gradual ramp, only a near-vertical
ascent to its high plateau}. When the cliff is encountered for the
first time by 16- or 17-year olds, it is small wonder that only a few
have courage (and the skill) to climb it. {\it There is no good reason
for this difference of intellectual topography.  First-graders could
be taught some physics}\dots (Hammer 1999)\dots, {\it second-graders a
little more, and third-graders still more}.  Then for the eleventh- or
twelfth-grader, a physics course would be a manageable step
upward. Some might choose to take it, some not, but few would be
barred by lack of `talent' or `background'.  [Our italics.]
\endquote

Forman, G. and P.B. Pufall, eds. (1988). \t{Constructivism in the
Computer Age}.  Erlbaum.

French, A.P. (1997). The Nature of Physics.  In Tiberghien, A.,
E.L.~Jossem, and J.~Barojas, eds., \t{Connecting Research in Physics 
Education with Teacher Education}, International Commission on Physics 
Education.  Online at 
\url{http://www.physics.ohio-state.edu/~jossem/ICPE/B1.html}.

Fullan, M.G. (1991). \t{The New Meaning of Educational Change}.
Teachers' College Press.

Fuller, R.G. , R. Karplus, and A.E. Lawson (1977). Can physics develop
reasoning?  \t{Physics Today}, February, pp.~23--28.

Fuller, R.G. (1982). Solving physics problems -- how do we do it? 
\t{Physics Today}, September, pp.~43--47.

Fuller, R.G. (1993).  Millikan lecture 1992; Hypermedia and the 
knowing of physics: Standing upon the shoulders of giants. 
\j{American Journal of Physics}{61\rm(4)}{300--304}.

Karplus, R. (2001). \t{A Love of Discovery: Science Education and the
Second Career of Robert Karplus}, R.G. Fuller, ed., to be published.

Gardner, H. (1985). \t{The Mind's New Science: A History of the
Cognitive Revolution}. Basic Books. See especially `Jean Piaget's
Developmental Concerns', pp. 116--118.

Gardner, H. (1985). \t{Frames of Mind: The Theory of Multiple
Intelligences}. Basic Books.

Gardner, H. (1991). \t{The Unschooled Mind: How Children Think \&\ How
Schools Should Teach}. Basic Books. Especially Chapter 2,
`Conceptualizing the Development of Mind'.

Gardner, H. (1999). \t{The Disciplined Mind: What All Students Should
Understand}. Simon \&\ Schuster.

Gardner, H. (2000). \t{Intelligence Reframed: Multiple Intelligences
for the 21st Century}. Basic Books.

D. Gentner and A.L. Stevens, eds. (1983). \t{Mental Models}.  Erlbaum.

Goldberg, F. and S. Bendall (1995). Making the invisible visible: A
teaching/learning environment that builds on a new view of the physics
learner.  \j{American Journal of Physics}{63}{978--991}.

Goodland, J.I. (1990). \t{Teachers for Our Nation's Schools}.
Jossey--Bass.

Goodland, J.I. (1994). \t{Educational Renewal: Better Teachers, Better
Schools}. Jossey-Bass.

Goodstein, D. (2000). The Coming Revolution in Physics Education.
\t{APS 
News}, June. Online at
\url{http://www.aps.org/apsnews/0600/060017.html}.

Haber-Schaim, U., J.H. Dodge, R. Gardner, E.A. Shore (1991). \t{PSSC
Physics}.  Kendall/Hunt. See Arons (1960) and Swartz (1999) for
reviews.

Haber-Schaim, U., R. Cutting, H.G. Kirksey, H. Pratt (1999).  
\t{Introductory Physical Science}.  7th ed.
Science Curriculum Inc.
\url{http://www.sci-ips.com/}.  See Raloff (2001) for a review; also 
mentioned but not reviewed by Hubisz et al. (2001).

Hake, R.R. (1987). Promoting student crossover to the Newtonian 
world. \j{American Journal of Physics}{55\rm(10)}{878--884}.

Hake, R.R.. (1992). Socratic pedagogy in the introductory physics
lab. \j{Physics Teacher}{30}{546--552}.  Version updated 27 April 1998,
online at \url{http://physics.indiana.edu/~sdi/}.

Hake, R.R. (1998a). Interactive-engagement vs traditional methods: A
six-thousand-student survey of mechanics test data for introductory
physics courses.  \ajp{66}{64--74}.
Online at \url{http://www.physics.indiana.edu/~sdi/}.

Hake, R.R. (1998b).  Interactive-engagement methods in introductory
mechanics courses.  Online at
\url{http://www.physics.indiana.edu/~sdi/} and submitted on 19 June
1998 to the \t{Physics Education Research Supplement to AJP (PERS)}.
See especially footnotes \#39 on the Socratic method, \#57 on
classroom communication systems, and \#66 on sources of conceptual
questions and problems.

Hake, R.R. (1999). \t{REsearch, Development, and Change in Undergraduate 
Biology Education: A Web Guide for Non-Biologists (REDCUBE)}.  Online 
at 
\url{http://www.physics.indiana.edu/~redcube}.

\*Hake, R.R. (2000a). Is it finally time to implement curriculum S? 
\j{AAPT 
Announcer}{30\rm(4)}{103}.  Online at
\url{http://www.physics.indiana.edu/~hake}.

\*Hake, R.R. (2000b). The general population's ignorancce of
science-related societal issues: A challenge for the university.
\j{AAPT Announcer}{30\rm(2)}{105}.  Online at
\url{http://physics.indiana.edu/~hake}.

Hake, R.R. (2001a). Lessons from the physics education reform effort.
Accepted for publication in \t{Conservation Ecology}.  Online at
\url{http://www.physics.indiana.edu/~hake}.

\*Hake, R.R. (2001b).  Suggestions for administering and reporting 
pre/post diagnostic tests. Unpublished.  Online at 
\url{http://physics.indiana.edu/~hake/}. 

Hammer, D. (1995). Epistemological considerations in teaching
introductory physics.  \j{Science Education}{79}{393--413}.

Hammer, D. (1997). Discovery learning and discovery
teaching. \j{Cognition and Instruc.}{15\rm(4)}{485--529}. Abstract
Online
at \url{http://www.physics.umd.edu/perg/cpt.html}. 

Hammer, D. (1999). Physics for first-graders? 
\j{Science Education}{83\rm(6)}{797--799}.
Online at  \url{http://www.physics.umd.edu/perg/cpt.html}.

Hammer, D. (2000). Student resources for learning introductory 
physics. \t{Physics Education Research}, supplement 1 to the
\ajp{68\rm(7)}{S52--S59}.

Hammer, D. (2001). Teacher inquiry. To appear in J. Minstrell and E. van Zee,
eds., \t{Teaching and Learning
in an Inquiry-Based Science Classrooom}.  (Also to appear in the Paper
Series of the Center for the Development of Teaching at EDC, in Newton, MA.)
Pre-publication draft
online at \url{http://www.physics.umd.edu/perg/cpt.html}. 

J. Handlesman, B. Houser, and H. Kriegel (1997). \t{Biology Brought to
Life: A guide to teaching students to think like scientists}. Times
Mirror Higher Education Group.

Hargan, J.D.D. and
M.S. Rivkin (1999). \t{Science Experiences in the Early Childhood Years:
An Integrated Approach}. Prentice--Hall. Recommended by Hubisz et
al. (2001). 

Harte, J. (1988).  \t{Consider a Spherical Cow: A Course in
Environmental Problem Solving}. University Science Books.

Hawkins, D. (1974a).  On Living in Trees.  In \t{The Informed Vision:
Essays on Learning and Human Nature\/} (Agathon Press).

Hawkins, D. (1974b). Messing about in science.  In \t{The Informed
Vision: Essays on Learning and Human Nature\/} (Agathon Press).

Hawkins, D. (1974c).  Critical barriers to science learning.  Online
at
\url{http://www.exploratorium.edu/IFI/resources/museumeducation/criticalbarriers.html}.

Hawkins, D. (2001). \t{The Roots of Literacy}. University of Colorado
Press.

Hazen R.M. and J. Trefil (1991). \t{Science Matters: Achieving Scientific 
Literacy}.  Doubleday.

Heibert, J., T.P. Carpenter, and E. Fennema (1997). \t{Making Sense:
Teaching and Learning Mathematics With Understanding}. Heinemann.

Heller, K.J. (1999). Introductory physics reform in the traditional 
format: an intellectual framework. \t{APS Forum on Education Newsletter},
Summer.  Online at\hb
\url{http://webs.csu.edu/~bisb2/FEdnl/heller.htm}.

Heller, K.J. (2001). The time has come to make teaching a real
profession.  \t{APS Forum on Education Newsletter}, Spring.  Online at
\url{http://www.aps.org/units/fed/index.html}.  Suggests raising K--12 
teacher's average salary so as to equal mechanical engineer's average
salary, with a Fermi-problem estimated cost of about 450 billion over
10 years. A similar proposal and cost estimate was made by Langenberg
(2000).

Herron J. D. and S.C. Nurrenbern (1999).  Chemical education research:
Improving chemistry learning. \j{Journal of Chemical
Education}{76\rm(10)}{1353--1361}.

Hestenes, D.  (1992).  Modeling games in the Newtonian world.  
\ajp{60\rm(8)}{732--748}.

Hillocks, G. (1999).  \t{Ways of Thinking, Ways of Teaching}.  Teachers' 
College Press.

Hobson, A. (2000/2001). Teaching relevant science for scientific
literacy. \t{Journal of College Science Teaching},
December 2000/January 2001.

Hobson, A. (1999).  \t{Physics: Concepts and
Connections}.  Prentice--Hall.

Holton, G. (1998/99). The New Imperative for Science Literacy.  
\t{Journal of 
College Science Teaching}, Dec. 1998/January 1999, 181--185.

Holton, G. and S.G. Brush. (2001). \t{Physics, the Human Adventure: From
Copernicus to Einstein and Beyond}. Rutgers University Press.

Hubisz, J.L. et al. (2000?).  \t{Review of Middle School Physical
Science Texts}. Online at
\url{http://www.science-house.org/middleschool/reviews/}.

Hubisz, J.L.  et al. (2001).  Report on a study of middle school 
science texts.  \j{Physics Teacher}{39\rm(5)}{304--309}.
\beginquote
If good materials 
are to be used, we must bring them to the attention of teachers and 
administrators.  Our two year search, unfortunately, has led us to 
say {\it that the available text books are not the tools that will effect 
a change in the way physical science is taught in the middle schools 
of the United States}.  [emphasis added]
\endquote
\hskip\a
With seeming inconsistency, 
Raloff (2001) reports that
\beginquote
\dots Hubisz\dots would have liked to include the book
\dots Haber-Schaim et al. (1999) -- an `available' textbook\dots
in his 
recent review of science texts `because it would have allowed us to 
say something really positive.' But since it was not among the top 
dozen sellers, it didn't make the cut.
\endquote
\hskip\a
See especially `Suggestions for Middle School Teachers'.

Hurd, P.D. (1994). New minds for a new age: Prologue to modernizing
the science curriculum. \j{Science Education}{78\rm(1)}{103--116}.

Hurd, P.D. (1997). \t{Inventing Science Education for the New Millennium}.
Teachers' College Press.

Hurd, P.D. (2000).  \t{Transforming Middle School Science}. Teachers
College Press. 

Inhelder, B., D. DeCaprona, and A. Cornu-Wells (1987).  \t{Piaget Today}.
Psychology Press.

Inhelder, B. and J. Piaget (1958). \t{Growth of Logical Thinking from 
Childhood to Adolescence: An Essay on the Construction of Formal 
Operational Structures}. Basic Books.

Jungck, J.R. (1972).  The Three I's: Interdisciplinary, 
Investigative, and Independent study.  In 
R.K. Gibbs and P.A. Taylor, eds., 
\t{How Ought Science Be Taught?}, MSS Educational 
Publishers, pp. 209--210.

Jungck, J.R. (1985).  A Problem Posing Approach to Biology Education.
\j{American Biology Teacher}{47\rm(5)}{264--266}.
Reprinted in S. Brown and M. Walter, eds., \t{Problem Posing:
Reflections and Applications}, Erlbaum, 1993.

Jungck, J.R. and N.S Peterson (1988). Problem-posing, problem-solving,
and persuasion in biology education.  \j{Academic Computing}{2\rm(6)}
{14--17, 48--50}.

Kaput, J.J. (1992).  Technology and mathematics education.  In
D.A. Grouws, ed., \t{Handbook of Research on Mathematics Teaching and
Learning}, Macmillan.

Karplus, R. (1977). Science teaching and the development of
reasoning. \j{Journal of Research in Science Education}{14}{169}.

Kessler, J.H., A. Benbrow, and T. Sweitzer (1996). \t{The
Best of Wonder Science: Elementary Science Activities}.  Delmar Thomson
Learning. Recommended by Hubicz et al. (2001). 

Kindfield, A.C.H. (1994). Assessing understanding of biological
processes: Elucidating students models of meiosis.  \j{American
Biology Teacher}{56\rm(6)}{367--371}.

King, J.G. (2001).  `Observation, experiment, and the future of
physics': John G. King's acceptance speech for the 2000 Oersted medal
presented by the American Association of Physics Teachers, 18 January
2000.  \ajp{69\rm(1)}{11--25}.  From the abstract:
\beginquote
Looking at our built world, most physicists see order where many
others see magic. This view of order should be available to all, and
physics would flourish better in an appreciative society. Despite the
remarkable developments in the teaching of physics in the last half
century, too many people, whether they've had physics courses or not,
don't have an inkling of the power and value of our subject, whose
importance ranges from the practical to the psychological. We need to
supplement people's experiences in ways that are applicable to
different groups, from physics majors to people without formal
education.  I will describe and explain an ambitious program to
stimulate scientific, engineering, and technological interest and
understanding through direct observation of a wide range of phenomena
and experimentation with them. For the very young: toys, playgrounds,
kits, projects. For older students: indoor showcases, projects, and
courses taught in intensive form. For all ages: more instructive
everyday surroundings with outdoor showcases and large demonstrations
\endquote

Kohn, A. (1999). \t{Punishment By Rewards: The Trouble with Gold
Stars, Incentive Plans, A's, Praise, and Other Bribes}. Houghton
Mifflin.

Kolb, D.A. (1984).  \t{Experiental Learning: Experience as the Source
of Learning and Development}. Prentice--Hall.

Langer, E.J. (1997). \t{The Power of Mindful Learning}.  Addison--Wesley.

Langenberg, D.N. (2000). Rising to the challenge. \j{Thinking K--16}
{4\rm(1)}{19}.  Online as `Honor in the Boxcar'
\url{http://www.edtrust.org/main/reports.asp}: 
\beginquote
\dots on average, 
teacher's salaries ought to be about 50\%\ higher than they are now. 
Some teachers, including the very best, those who teach in shortage 
fields (e.g., math and science) and those who teach in the most 
challenging environments (e.g., inner cities) ought to have salaries 
about twice the current norm\dots Simple arithmetic applied to 
publicly available data shows that the increased cost would be only 
0.6\%\ of the GDP, about one-twentieth of what we pay for health care. 
I'd assert that if we can't bring
ourselves to pony up that amount,
we will pay far more dearly in the long run.
\endquote

Lave J. and E. Wenger (1991).  \t{Situated Learning: Legitimate
Peripheral Participation}. Cambridge University Press.

Laws, P. (1997). Millikan lecture 1996: Promoting active learning 
based on physics education research in introductory physics courses.
\ajp{65\rm(1)}{13--21}.

Lawson, A.E. (1993a).  Deductive reasoning, brain maturation, and
science concept acquisition: Are they linked? \j{Journal of Research
in Science Teaching}{30\rm(9)}{1029--1052}.

Lawson, A.E. (1993b). Neural principles of memory and a neural theory
of analogical insight.  \j{Journal of Research in Science
Teaching}{30\rm(10)}{1327--1348}.

Lawson, A.E. ed. (1993c).  Special issue on the role of analogy in
science and science teaching.  \j{Journal of Research in Science
Teaching}{30\rm(10)}{1211--1364}.

Lawson, A.E., W.P. Baker, L. DiDonato, M.P. Verdi, and M.A. Johnson
(1993). The role of physical analogues of molecular interactions and
hypothetico-deductive reasoning in conceptual change. \j{Journal of
Research in Science Teaching}{30\rm(9)}{1073--1086}.

Lawson, A.E. (1994). Epistemological foundations of cognition.  In D.
Gabel, ed., \t{Handbook of Research on Science Teaching and Learning},
Macmillan.

Lawson, A.E. (1995). \t{Science Teaching and the Development of
Thinking}.  Wadsworth Publishing.  See especially Appendix C, `What is
Science' by R.P. Feynman; Appendix F, `Classroom Test of Scientific
Reasoning'; and the excellent bibliography.

Lawson, A.E. (1999). What should students learn about the nature of 
science and how should we teach it? \j{Journal of College Science 
Teaching}{28\rm(6)}{401}.

Layman, J. W., G. Ochoa, and H. Heikkinne (1996).  \t{Inquiry and 
Learning: Realizing Science Standards in the Classroom}.  College 
Entrance Examination Board.  Online at 
\url{http://www.nap.edu/readingroom/books/rtmss/7.19.html}.

\*Lederman, L.M. (1999). A science way of thinking.  \t{Education Week}, 16
June.  \url{http://www.edweek.org/ew/1999/40leder.h18}.

\*Livanis, O. (2000).  \t{Physics First Home Page}.
\url{http://members.aol.com/physicsfirst/index.html}.

Lochhead, J. (2000). \t{Thinkback: A User's Guide to Minding the
Mind}.  Erlbaum.  See also at
\url{http://www.whimbey.com/Books/Thinkback/thinkback.htm}

Ma, L. (1999). \t{Knowing and Teaching Elementary Mathematics:
Teachers' Understanding of Fundamental Mathematics in China and
theUnited States}. Erlbaum. For a review of see R. Askey, \t{American
Educator}, Fall 1999; on the web at the back-to-basics site
`Mathematically Correct'
\url{http://ourworld.compuserve.com/homepages/mathman/}.

Marshall R. and M. Tucker. (1992).  \t{Thinking for a Living}. Basic
Books.

McDermott, L.C. (1990).  A perspective on teacher preparation in 
physics and other sciences: The need for special science courses for 
teachers. \ajp{58}{734--742}.

McDermott, L.C. (1991).  Millikan lecture 1990: What we teach and 
what is learned: Closing the gap. \ajp{59\rm(4)}{301--315}.

McDermott, L.C. (1993). Guest comment: How we teach and how students 
learn -- a mismatch? \ajp{61\rm(4)}{295--298}.

McDermott, L.C. (2000). Oersted medal lecture: Research -- key to
student understanding. \j{AAPT Announcer}{30\rm(4)}{88}.

McDermott L.C. and Redish E.F. (1999).  RL-PER1: Resource letter on 
physics education research. \ajp{67\rm(9)}{755--767}.  Online at
\url{http://www.physics.umd.edu/rgroups/ripe/perg/cpt.html}.

Mervis, J. (2001).  Math education: Academy report aims to quiet
debate.  \j{Science}{291}{808} (2 February).  Describes a panel report
by the National Academy of Sciences
\url{http://www.nap.edu/catalog/9822.html} on the `skills versus
understanding' wars in mathematics education.  The article quotes the
chair of the panel, Professor Jeremy Kilpatrick of the University of
Georgia:
\beginquote
We want to move past the debate over skills versus understanding. It's
not one or the other.  The point is that both are needed, and more, to
learn and understand mathematics.
\endquote

Mestre, J.P. (1991). learning and instruction in pre-college physical
science. \j{Physics Today}{44\rm(9)}{56--62}.  \beginquote Traditional
teaching practices only poorly reflect what is known about the
learning process. To improve science education, teachers and
scientists must take note of the implications of cognitive science.
\endquote

Mestre, J. and J. Touger (1989). Cognitive research -- what's in it for
physics teachers? \j{Physics Teacher}{27}{447--456}.

Minstrell, J. (1989). Teaching science for understanding. In L.B.
Resnick and L.E. Klopfer, eds., \t{Toward the Thinking Curriculum:
Current Cognitive Research}. Association for Supervision and
Curriculum Development [ASCD] Yearbook.

Moore, J.A. (1993). We need a revolution -- teaching biology for the 
twenty-first century. \j{Bioscience}{43\rm(11)}{782--786}.

Moore, J.A. (1993). \t{Science as a Way of Knowing: The Foundations of 
Modern Biology}.  Harvard University Press.

Moyer, A.E. (1982). John Dewey on physics teaching.  
\pt{20\rm(3)}{173--175}.

\*NISE [National Institute for Science Education] (2001).
\url{http://www.wcer.wisc.edu/NISE}: Field-tested Learning Assessment 
Guide (FLAG): For Science, Math, and Engineering Instructors.
\url{http://www.wcer.wisc.edu/nise/cl1/flag/flaghome.asp}, especially
Introduction to CAT's [Classroom Assessment Techniques] at
\url{http://www.wcer.wisc.edu/nise/cl1/flag/cat/catframe.asp}.

National Research Council (1989).  \t{Everybody Counts: A Report to the
Nation on the Future of Mathematics Education}. Mathematical Sciences
Education Board and the Board on Mathematical Sciences.  Online at
\url{http://www.nap.edu/catalog/1199.html}.

National Research Council (1990). \t{Fulfilling the Promise: Biology
Education in the Nation's Schools}. National Academy Press.  For a
description see\hb
\url{http://www.nap.edu/bookstore/isbn/0309051479.html}.

\*National Research Council (1996a). \t{National Science Education
Standards}. National Academy Press.
\url{http://books.nap.edu/catalog/4962.html}.

National Research Council (1996b). 
\t{Resources for Teaching Elementary School Science}.
National Academy Press.  Online at
\url{http://www.nap.edu/catalog/4966.html}.

National Research Council (1997a). \t{Introducing the National Science
Education Standards}, Booklet. National Academy Press. Online at
\url{http://books.nap.edu/catalog/5704.html}.

National Research Council (1997b). \t{Improving Teacher Preparation and
Credentialing Consistent with the National Science Education
Standards: Report of a Symposium}. National Academy Press. Online at
\url{http://books.nap.edu/catalog/5592.html}.

National Research Council (1997c). \t{Science Teaching Reconsidered: A
Handbook}. National Academy Press.  Online at
\url{http://books.nap.edu/catalog/5287.html}.

National Research Council (1997d). \t{Science for All Children: A Guide
to Improving Elementary Science Education in Your School District}.
National Academy Press.  Online at
\url{http://books.nap.edu/catalog/4964.html}.

National Research Council (1997e). \t{Preparing for the 21st Century: 
The Education Imperative}. National Academy Press. Online at 
\url{http://books.nap.edu/catalog/9537.html}.

\def\nap{National Academy Press}

National Research Council (1998a). \t{Resources for Teaching Middle
School Science}. \nap.  Online at
\url{http://books.nap.edu/catalog/5774.html}.

National Research Council (1998b). \t{Every Child a Scientist: Achieving 
Scientific Literacy for All}.  \nap.  Online at 
\url{http://books.nap.edu/catalog/6005.html}.

National Research Council (1999a).  \t{Global Perspectives for Local 
Action: Using TIMSS to Improve U.S. Mathematics and Science Education}.
\nap. Online at
\url{http://www.nap.edu/catalog/9605.html}.

National Research Council (1999b). \t{Improving Student Learning: A 
Strategic Plan for Education Research and Its Utilization}. \nap.  Online at
\url{http://books.nap.edu/catalog/6488.html}.

National Research Council (1999c). \t{Selecting Instructional Materials: 
A Guide for K--12 Science}. \nap. Online at 
\url{http://books.nap.edu/catalog/9607.html}.

National Research Council (1999d). \t{Designing Mathematics or Science 
Curriculum Programs: A Guide for Using Mathematics and Science 
Education Standards}. \nap.  Online at
\url{http://books.nap.edu/catalog/9658.html}.

National Research Council (2000a). \t{Educating Teachers of Science, 
Mathematics, and Technology: New Practices for the New Millennium}. 
\nap.  Online at
\url{http://www.nap.edu/catalog/9832.html}.

National Research Council (2000b). Steve Olson and Susan
Loucks-Horsley, eds. Committee on the Development of an Addendum
to the National Science Education Standards on Scientific Inquiry,
National Research Council.
\t{Inquiry and the `National Science Education Standards': A Guide for 
Teaching and Learning}. \nap.  Online at 
\url{http://books.nap.edu/catalog/9596.html}.

National Research Council (2001a). Committee on Development of an 
Addendum to the National Science Education Standards on Science and 
Technology.  \t{Science and Technology and the National Science Education 
Standards: A Guide for Teaching and Learning}. \nap.
Final forthcoming / no prepublication copies for sale; see
\url{http://books.nap.edu/catalog/9833.html}.

National Research Council (2001b). J. Myron Atkin, Paul Black, and
Janet Coffey, eds. \t{Classroom Assessment and the National Science
Education Standards: A Guide for Teaching and Learning}.
\nap.  Final forthcoming / no prepublication copies for sale; see
\url{http://books.nap.edu/catalog/9847.html}.

National Research Council. (2001c). Committee on Understanding the 
Influence of Standards in Science, Mathematics, and Technology 
Education, Iris R. Weiss, Karen S. Hollweg, and Gail Burrill, eds.
\t{Considering the Influence: A Framework for Research on the Effects of 
Nationally Developed Mathematics, Science, and Technology Education 
Standards}. \nap. Final forthcoming / no prepublication 
copies for sale; see
\url{http://books.nap.edu/catalog/10023.html}.

National Science Foundation (no date).  \t{Inquiry: Thoughts, Views,
and Strategies for the K--5 Classroom}.  Online at
\url{http://www.nsf.gov/pubs/2000/nsf99148/htmstart.htm}.


Nelson, C.E. (1989).  Skewered on the unicorn's horn: The illusion of
tragic tradeoff between content and critical thinking in the teaching
of science.  In L.W. Crow, ed.,
\t{Enhancing Critical Thinking in the Sciences},  Society for College
Science Teachers/National Science Teachers Association, pp. 17--27.

Nelson, C.E. (1994). Critical thinking and collaborative learning.
\j{New Directions for Teaching and Learning}{59}{45--58}.

Nelson, C. (2000). Bibliography: how to find out more about college
teaching and its scholarship: A not too brief, very selective
hyperlinked list; college pedagogy {\it is\/} a major area of
scholarship!.
\url{http://php.indiana.edu/~nelson1/TCHNGBKS.html}.

Neuschatz, M. (1999). What can the TIMSS teach us? \j{Science 
Teacher}{66\rm(1)}{23--26}.

\*NCSU [North Carolina State University] (2001).  Physics Education R
\&\ D Group, Assessment Instrument Information Page, online at
\url{http://www.ncsu.edu/per/TestInfo.html}.

Novak, G. M., E.T. Patterson, A.D. Gavrin, and W. Christian. (1999). \t{Just 
in Time Teaching}. Prentice--Hall.  For a description see 
\url{http://webphysics.iupui.edu/jitt/jitt.html}.

Novak, J.D., (1998). \t{Learning, Creating, and Using Knowledge: Concept 
Maps as Facilitative Tools in Schools and Classrooms}. Erlbaum.

Novak, J.D., J.J. Mintzes, and J.H. Wandersee,
eds. (1998). \t{Teaching Science for Understanding -- A Human
Constructivist View}. Academic Press.

Nuffield Foundation (1999).  \t{Beyond 2000 -- Science Education for
the Future}. Nuffield Foundation, UK.  Produced by prominent UK
science educators; makes 10 major recommendations for the future
direction of science education.  Online at\hb
\url{http://www.kcl.ac.uk/depsta/education/be2000/index.html}.

Paulos, J.A. (1992). \t{Beyond Numeracy: Ruminations of a Numbers
Man}.  Vintage Books.

Perkins, D. (1992). \t{Smart Schools: Better Thinking and Learning for 
Every Child}. Free Press.

Perry, W.G. (1970).  \t{Forms of Intellectual and Ethical Development 
in the College Years: A Scheme}. Holt, Rinehart, and Winston.

Pinker, S. (1999).  \t{How the Mind Works}. Norton. See also
\url{http://web.mit.edu/bcs/pinker.html}.

Polya, G. (1954).  \t{Mathematics and Plausible Reasoning\/} (2
volumes).  Volume I:
\t{Induction and Analogy in Mathematics}.  Volume II: \t{Patterns of
Plausible Inference}.  Princeton University Press.  At a higher level
than Polya (1957); his most brilliant books on mathematics teaching
and thinking.

Polya, G. (1957). \t{How to Solve It: A New Aspect of Mathematical
Method}. Princeton University Press.

Polya (1962, 1965).  \t{Mathematical Discovery: On Understanding,
Learning, and Teaching Problem Solving\/} (2 volumes).  Wiley.  Almost
as good as Polya (1954) and contains his famous `Ten Commandments for
Teachers' (p. 116).

Polya, G. (1965). \t{Let Us Teach Guessing\/} (60-minute videotape).
Mathematical Association of America.  Available from
\url{http://www.maa.org/pubs/books/ltg.html}.

\*Radatz, H. (1983).  Untersuchungen zum L\"osen eingekleideter
Aufgaben.  [Studies of solving mathematical tasks] \j{Journal f\"ur
Mathematikdidaktik}{3}{205--217}.

Raloff, J. (2001). Where's the book? Science education is redefining 
texts. \t{Science News\/} {\bf 159\rm(12)} (March 24).  Online at 
\url{http://www.sciencenews.org/20010324/bob12.asp} (second in a two-part 
series on middle-school science curricula). This article extols 
Haber-Schaim et al. (1999): 
\beginquote
Common to the origin of many of these\dots
(newer science education programs)\dots
is funding from the NSF. 
Explains Janice Earle, a senior program director\dots(at NSF)\dots
qualifying projects must now exhibit `a coherent content\dots
aligned with national standards', foster critical thinking and 
problem solving, and be grounded in research on how children learn. 
Moreover, NSF recommends that any new curriculum be developed by 
teams of practicing scientists, engineers, and mathematicians, along 
with classroom teachers.  `I would be surprised if most  textbooks 
were developed like that,' Earle says.  They aren't. One exception, 
however, is (Haber-Schaim et al. 1999)\dots notes Uri 
Haber-Schaim, one of this textbook's authors.  Launched in 1967, the 
book briefly became a top selection for eighth- and ninth-grade 
classrooms. Developed with NSF funding, the book was initially issued 
by a big publisher, but sales dropped when newer texts entered the 
field. In the early 1990's, the company decided not to publish further 
editions but permitted Haber-Schaim to pick up rights to the book. 
His firm, Science Curriculum Inc. of Belmont, Mass., now produces it. 
Unlike other science texts for early adolescents, Haber-Schaim says, 
`we very thoroughly field-tested our experiments in classrooms
over a period of 2 years. We even field-tested every homework 
question\dots(Among extollers)\dots of the book is John L. Hubisz 
\dots He would have liked to include the book in his recent review of 
science texts\dots \url{http://www.aapt.org} `because it would have 
allowed us to say something really positive.' But since it was not 
among the top dozen sellers, it didn't make the cut.
\endquote

Rath, A. and D. E. Brown (1996). Modes of engagement in science 
inquiry: A microanalysis of elementary students' orientations toward 
phenomena at a summer science camp. \j{Journal of Research in Science 
Teaching}{33}{1083--1097}.

Redish, E.F. (1994). Implications of cognitive studies for teaching
physics.  \ajp{62\rm(9)}{796--803}.  Online at
\url{http://www.physics.umd.edu/rgroups/ripe/perg/cpt.html}.

Redish, E.F. (1999). Millikan lecture 1998: Building a science of 
teaching physics.  \ajp{67\rm(7)}{562--573}.  Online at
\url{http://www.physics.umd.edu/rgroups/ripe/perg/cpt.html}.

Reif, F. (1995). Millikan lecture 1994: Understanding and teaching 
important scientific thought processes.  \ajp{63\rm(1)}{17--32}.

Reif, F. (1999). Thermal physics in the introductory physics course: 
Why and how to teach it from a unified atomic perspective. 
\ajp{67\rm(12)}{1051--1062}.

Resnick, L.B. and M.W. Hall (1998).  Learning organizations for
sustainable education reform. \j{Daedalus}{127\rm(4)}{89--118}.

\*Reusser, Kurt (1988).  Problem Solving
beyond the Logic of Things: Contextual Effects on Understanding and
Solving Word Problems.  \j{Instructional Science}{17}{309--38}.

Sadler P.M. and R.H. Tai. (1997). The role of high-school physics in 
preparing students for college physics.  \pt{35\rm(5)}{282--285}.

Salinger, G.L. (1991).  The materials of physics instruction.
\j{Physics Today}{44\rm(9)}{39--45}.  \beginquote The mathematics and
science communities are redefining what students should know and how
they should learn it. Much new instructional material is available
embodying the new reforms.  \endquote

Sarason, S.B. (1993). \t{The Case for Change: Rethinking the Preparation 
of Educators}. Jossey--Bass.

Sarason, S.B. (1996).  \t{Revisiting `The culture of the school and the 
problem of change'}. Teachers' College Press.

Schneps, M.H. and P.M. Sadler (1985).  Private universe project.
Harvard--Smithsonian Center for Astrophysics, Science Education 
Department.\hb
\url{http://cfa-www.harvard.edu/cfa/sed/resources/privateuniv.html}.

Schwartz, R.B. (1999).  Lessons from TIMSS.  \j{Hands On}{21\rm(1)}{4}.
\rawurl{http://www.terc.edu/handsonIssues/s98/schwartz.html}
{http://www.terc.edu/handson\1Issues/s98/schwartz.html}.
\t{Hands On\/} is a publication of TERC [Technical Education Research Center]
\url{http://www.terc.edu}.  Schwartz is president of ACHIEVE (2001), and 
on the faculty of the Harvard Graduate School of Education.

Shapiro, I., C. Whitney, P. Sadler, and M. Schneps (1997).  \t{Simple
Minds}.
Harvard--Smithsonian Center for Astrophysics, Science Education 
Department, Media Group.
\url{http://www.learner.org/catalog/science/mooo/} and
\url{http://www.learner.org/catalog/science/mooo/mooodes.html}.

Simon, H.A. (1998).  What we know about learning.  \j{Journal of 
Engineering Education}{87\rm(4)}{343--348}.

Schoenfeld, A.H. (1985).  \t{Mathematical Problem Solving}. Academic Press.

\*Schoenfeld, Alan (1987).  What's all the fuss about metacognition? In
\t{Cognitive Science and Mathematics Education}, Alan
Schoenfeld, ed. (Hillsdale, NJ: Lawrence Erlbaum).

\*Schoenfeld, Alan H. (1989).  Teaching mathematical thinking and
problem solving.  In Lauren B. Resnick and Leonard B. Klopfer, eds.,
\t{Toward the thinking curriculum: Current cognitive research},
Association for Supervision and Curriculum Development [ASCD],
Alexandria, VA.

Schultz, T. (2001).  \t{Science Education Through the Eyes of a 
Physicist}. National Academy Of Sciences/National Research Council.
Online at \url{http://www.nas.edu/rise/backg2d.htm}.

\def\HM{Houghton Mifflin}
Sizer, T.R. (1985.)  \t{Horace's Compromise: The Dilemma of the
American High School}. \HM.

Sizer, T.R. (1992). \t{Horace's School: Redesigning the American High 
School}. \HM.

Sizer, T.R. (1996).  \t{Horace's Hope: What Works for the American High 
School}. \HM.

Sizer, T.R. and N.F. Sizer (1999).  \t{The Students Are Watching: Schools 
and the Moral Contract}. Beacon Press. See also 
\url{http://www.essentialschools.org/}.

Slavin, R.E. (1994). \t{Cooperative Learning: Theory, Research and 
Practice}. Allyn \&\ Bacon.

Springer, L., M.E. Stanne, and S.D. Donovan (1999). Undergraduates 
in science, mathematics, engineering, and technology: a 
meta-analysis.  \j{Review of Educational Research}{69\rm(1)}{21--51}.
Abstract online at \url{http://www.aera.net/pubs/rer/abs/rer691-3.htm}.

Stein, S.K. (1996). \t{Strength in Numbers: Discovering the Joy and 
Power of Mathematics in Everyday Life}. John Wiley.

Sternberg, R.J. (1995). \t{In Search of the Human Mind}. Harcourt Brace.
Especially `Evaluation of Piaget's Theory', pp. 427--428.

Stevenson H.W., and J.W. Stigler (1992). \t{The Learning Gap: Why Our
Schools are Failing and What We Can Learn from Japanese and Chinese
Education}. Summit Books.

Stigler, J.W. and J. Hiebert (1997). Understanding and Improving 
Classroom Mathematics Instruction: An Overview of the TIMSS Video 
Study.  \t{Phi Delta Kappan}, September.
Online at \url{http://www.pdkintl.org/kappan/kstg9709.htm}.

Stigler, J.W. and J. Hiebert (1999). \t{The Teaching Gap}. 
Free Press.  See also
\url{http://www.lessonlab.com/teaching-gap/}.  Contains an illuminating 
analysis of TIMSS videotapes.

Swartz, C.E. (1969). Measure and find out: A quantitative approach to
science.  Scott, Foresman, \&\ Company.  3 volumes. Recommended by Hubisz
et al. (2001): `Has what is missing from most of the books in this
report'.

Swartz, C.E. (1986).  Elementary-school science by a quantitative
approach.  Unnpublished.

Swartz, C.E. (1991). The physicists
intervene. \j{Physics Today}{44\rm(9)}{22--28}.
\beginquote
For over 150 years American
physicists have been making forays into elementary and high school
science teaching. Their novel approaches have usually worked--but the
results have always been short-lived.
\endquote

\*Swartz, C.E. (1993). Editorial: Standard reaction. \j{Physics Teacher}{31}
{334--335}.

Swartz, C.E. et al. (1999). Survey of high-school texts.  
\pt{37\rm(5)}{283--296}.

Swartz C.E. and T. Miner (1998). \t{Teaching Introductory Physics: A 
Source Book}.  Springer-Verlag.  For a description see
\url{http://www.springer-ny.com/catalog/np/apr98np/1-56396-320-5.html}.

Schwartz, B.B. and J.J. Wynne (1991). Pre-college physics education
programs from the research community. \j{Physics
Today}{44\rm(9)}{48--54}.  \beginquote Physicists from professional
societies, the national laboratories and industry run many active
programs for teachers and students of pre-college physics.  \endquote

Tyack, D. and L. Cuban (1995).  \t{Tinkering Toward Utopia: A Century of 
Public School Reform}. Harvard University Press.

Uno, G.E. (1994) (principal author).  \t{Developing Biological
Literacy}.  Biological Sciences Curriculum Study [BSCS], Kendall/Hunt.

Uno, G.E. and R.W. Bybee (1994).  Understanding the dimensions of 
biological literacy.  \j{BioScience}{44\rm(8)}{553--557}.

Uno, G.E. (1992). \t{Co-author and editor of Biological Science: An 
Ecological Approach\/} (BSCS text). 7th edition.  Kendall/Hunt.

Uno, G.E. (1984). \t{Investigating the Human Environment: Land Use\/} (BSCS 
text). Kendall/Hunt.

Uno, G.E. (1988). Teaching College and College-Bound Biology 
Students.  \j{American Biology Teacher}{50\rm(4)}{213--216}.

von Glasersfeld, E. (1997).  Homage to Jean Piaget (1896-1982).
\j{Irish Journal of Psychology}{18\rm(2)}{293--}.  Online at 
\url{http://www.oikos.org/Piagethom.htm} and maybe at
\url{http://www.umass.edu/srri/vonGlasersfeld/onlinePapers.html}.

Wells, M., D. Hestenes, and G. Swackhamer (1995). A modeling method for 
high school physics instruction.  \ajp{63\rm(7)}{606--619}. Online at
\url{http://modeling.la.asu.edu/modeling/MalcolmMeth.html}.

West, L.H.T. and A.L. Pines, eds. (1985). \t{Cognitive Structure and 
Conceptual Change}. Academic Press.

Whimbey, A. and J. Lochhead (1999). \t{Problem Solving and
 Comprehension}.  6th edition.  Erlbaum.  See also at
\url{http://www.whimbey.com/Books/psc6/psc6.htm}.

Whimbey, A., J. Lochhead, and P.B. Potter (1999). \t{Thinking Through
Math Word Problems: Strategies for Intermediate Elementary School
Students}.  Erlbaum.  See also at
\url{http://www.whimbey.com/Books/thinking/thinking.htm}.

Williams, P.H. (1993). \t{Bottle Biology}. Kendall/Hunt.

Whitehead, A.N. (1967). \t{The Aims of Education, and Other Essays}.
Free Press. Originally published in 1929.

Wiggins, G. and J. McTighe (2005).  \t{Understanding by Design}.  2nd,
expanded edition.  Association for Supervision and Curriculum
Development [ASCD].  See also \url{http://authenticeducation.org}.

Wilson K.G. and B. Daviss (1994). \t{Redesigning Education}.  Henry Holt.
See also at
\url{http://www-physics.mps.ohio-state.edu/~kgw/RE.html}.

Wyckoff, S. (2001). Changing the culture of undergraduate science
teaching. \j{Journal of College Science
Teaching}{30\rm(5)}{306--312}. Figure~2 shows pre/post
Lawson-reasoning test [Lawson(1995/99)] results for 584 students in a
`Biology 100' class that `demonstrates' a 41\%\ gain in students
reasoning abilities. For related work see Clement (2000) and Adey
(1999).

Zollman, D.A. (1996). Millikan lecture 1995: Do they just sit there? 
Reflections on helping students learn physics. \ajp{64\rm(2)}{114--119}.

\endgroup

\end